\documentclass[aps,twocolumn,prd,superscriptaddress,nofootinbib,floatfix]{revtex4-2}

\usepackage{mathtools}
\usepackage{amsfonts}
\usepackage{amssymb}
\usepackage{mathrsfs}
\usepackage{bbm}
\usepackage{slashed}
\usepackage{amsmath}
\usepackage{bm}

\usepackage{graphicx}
\usepackage{color}
\usepackage{colortbl}
\usepackage{array}

\usepackage{float}
\usepackage{placeins}
\usepackage{booktabs}
\usepackage[caption=false]{subfig}
\usepackage{makecell}
\usepackage{tabstackengine}

\usepackage{xspace}
\usepackage{hyperref}
\usepackage[nameinlink]{cleveref}
\usepackage{bookmark}
\usepackage{siunitx}

\usepackage{xifthen}
\usepackage{xcolor}
\hypersetup{
	colorlinks,
	linkcolor={red!75!black},
	citecolor={blue!75!black},
	urlcolor={blue!75!black}
}

\usepackage[utf8]{inputenc}
\allowdisplaybreaks[4]

\newcommand{\feyn}[1]{
	\setbox0=\hbox{\ensuremath{#1}}
	\hbox to\wd0{\hbox to0pt{\hbox to\wd0{\hss/\hss}\hss}\box0}}

\setkeys{Gin}{width=0.48\textwidth}
\graphicspath{{./figures/}}

\newcommand{\gettitle}{Binding Energy}

\newcommand{\getHeidelbergAffiliation}{\affiliation{Institut f{\"u}r Theoretische Physik, Universit{\"a}t Heidelberg, Philosophenweg 16, 69120 Heidelberg, Germany}}

\newcommand{\getDalianAFfiliation}{\affiliation{School of Physics, Dalian University of Technology, Dalian, 116024, P.R. China}}

\hypersetup{
	colorlinks,
	linkcolor={red!75!black},
	citecolor={blue!75!black},
	urlcolor={blue!75!black},
	pdftitle={\gettitle},
	pdfauthor={Zelle},
	pdfkeywords={effective theory} {analytic continuation}
	{correlations functions} {hadronization}
	{functional renormalisation group}
	{real time} {bound states}
	bookmarksopen=true,
	bookmarksopenlevel=2,
	bookmarksnumbered=true
}

\begin{document}

\title{Quasi parton distributions of pions at large longitudinal momentum}
	
\author{Dao-yu Zhang}\getDalianAFfiliation

\author{Chuang Huang}\email{huang@thphys.uni-heidelberg.de}\getHeidelbergAffiliation
        
\author{Wei-jie Fu}\getDalianAFfiliation

\begin{abstract} 

In this paper, we develop an approach to calculate the valence-quark quasi parton distribution amplitude (quasi-PDA) and quasi parton distribution function (quasi-PDF) for the pion with a large longitudinal momentum with the functional renormalization group (fRG). This is demonstrated in a low energy effective theory (LEFT) with four-quark scatterings. In the study of the complex structure of quasi-PDA, we introduce a deformed integration contour in the calculations of quasi-PDA or quasi-PDF, which allows us to obtain correct integrals for all momentum fractions. It is found that the pion light-front PDA extrapolated from quasi-PDA based on the large momentum effective theory (LaMET) in the LEFT is comparable with lattice QCD and Dyson-Schwinger equation. This work paves the way to study the PDA and PDF within the fRG approach to first-principles QCD.

\end{abstract}
	
\maketitle

\section{Introduction}

The Standard Model of particle physics describes all known fundamental interactions and elementary particles. For the mass of visible matter, only 1-2\% originates from the Higgs mechanism in the Standard Model, while approximately 98\% comes from the strong interactions described by Quantum Chromodynamics (QCD), see \cite{Marciano:1977su, Marciano:1979wa}. In particular, the $\pi$ meson, as the Goldstone boson of dynamical chiral symmetry breaking in QCD, plays a fundamental role in both experimental and theoretical studies of QCD and hadron physics. 

The pion parton distribution amplitude (PDA) and parton distribution function (PDF) are the leading-twist physical quantities. The PDA describes the momentum distribution amplitude of the valence quark-antiquark inside the hadron, which is, e.g., a key input for the studies of relevant $B$ meson weak decay processes in the Standard Model and beyonds, see \cite{Cheng:2009cn, Su:2010vt, LHCb:2019hip}. The PDF provides the momentum distribution probabilities of partons in hadrons, including that of gluons, valence quarks and sea quarks. More importantly, the pion and kaon PDFs are key observables for studying hadron internal structures through deep inelastic scattering experiments \cite{Berger:1979du, Saclay-CERN-CollegedeFrance-EcolePoly-Orsay:1980fhh, NA3:1983ejh}. Measurement of the pion and kaon PDFs is also one of the main scientific goals in future electron-ion collider experiments \cite{Aguilar:2019teb, Anderle:2021wcy, Abir:2023fpo, Achenbach:2023pba}.

The direct computation of pion PDA or PDF requires to be done at the light cone, which usually is very challenging and difficult. They arise from the inherently complex structure of these observables, and one needs correlation functions in the complex plane as input to calculate them. Even though, significant progress was made in the direct calculation of PDA and PDF within effective model \cite{Eichmann:2021vnj}. However, there is still a long way to go before a realistic calculation in QCD is feasible.

Alternatively, we can also adopt indirect methods to compute the dependence of PDA or PDF on the momentum fraction of partons, usually denoted by $x$. One indirect method is to reconstruct the PDA or PDF from their Mellin moments or cumulants of different orders. Cumulants of several lowest-orders have already been calculated with high precision in lattice QCD \cite{Arthur:2010xf, Braun:2015axa, Bali:2017ude, RQCD:2019osh, Loffler:2021afv}, while for high-order ones the errors increase significantly. As a complement, cumulants of PDA or PDF can be calculated up to very high orders in the approach of Dyson-Schwinger equations (DSE), see e.g., \cite{Chang:2013pq, Chang:2014lva,  Chen:2016sno, Ding:2019lwe, Cui:2020tdf, Wang:2024fjt} and reviews \cite{Horn:2016rip, Yu:2024ovn}. Another important indirect method stems from the breakthrough in the study of relations between the correlation functions on the light-cone and those in the large momentum limit \cite{Ji:2013dva, Ji:2014gla} and the subsequent establishment of the Large Momentum Effective Theory (LaMET), see \cite{Ji:2020ect} for review. The LaMET allows us to extract the light-cone PDA or PDF from the Euclidean quasi parton distribution amplitude (quasi-PDA) or quasi parton distribution function (quasi-PDF) with a finite large longitudinal momentum by extrapolating to the limit of momentum infinity. The quasi-PDA method has already been applied in lattice QCD to compute the light-front wave function and PDA for the pion and kaon \cite{Zhang:2017bzy,LatticeParton:2022zqc,Holligan:2023rex,LatticeParton:2023xdl}.

In this work, we would like to introduce a novel approach to compute the pion quasi-PDA or quasi-PDF with the functional renormalization group (fRG). The fRG is a nonperturbative continuum field approach, which is a powerful theoretical method to study the nonperturbative QCD both in the vacuum and at finite temperature and densities \cite{Mitter:2014wpa, Braun:2014ata, Rennecke:2015eba, Cyrol:2016tym, Cyrol:2017ewj, Corell:2018yil, Fu:2019hdw, Braun:2020ada, Braun:2023qak, Ihssen:2024miv, Fu:2024rto}, see \cite{Dupuis:2020fhh, Fu:2022gou} for recent reviews. Notably, recently an approach of multi-quark flows within the fRG has been developed \cite{Fu:2022uow, Fu:2024ysj,Fu:2025hcm}, which is well suited for the studies of bound states, and also pave the necessary way for the studies of quasi parton distributions within the functional renormalization group. This is the central concern of this work.

This paper is organized as follows: Quasi parton distributions at large longitudinal momentum, relevant definitions and notations are briefly discussed in \Cref{sec:quasi-frame}. In \Cref{sec:fRG-input} we briefly discuss the approach of multi-quark flows within the fRG and show how the properties of bound states are extracted from the four-quark vertices. Analytic structure of the Bethe-Salpeter (BS) amplitude and quark mass in the complex plane is investigated in detail in \Cref{sec:Anal-struc} and numerical results are presented in \Cref{sec:numerical}. We conclude in \Cref{sec:conclusion}.

\section{Quasi parton distributions at large longitudinal momentum}
\label{sec:quasi-frame} 

In this Section, we discuss the quasi-light-front framework, which is used here to calculate the valence-quark parton distribution amplitude (PDA) and the parton distribution function (PDF) for the pions. This formalism is based on the Large Momentum Effective Theory (LaMET) \cite{Ji:2013dva, Ji:2014gla}, see also \cite{Ji:2020ect}. We will show that in the limit of large longitudinal momentum, the quasi-PDA returns to the analytical results of PDA obtained from the light-front calculations by using the pointlike and asymptotic cases as examples.

\begin{figure}[t]
\begin{align*}
\parbox[c]{0.2\textwidth}{\includegraphics[width=0.2\textwidth]{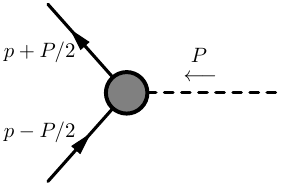}}=\chi_{\pi} (p;P)
\end{align*}
\caption{Diagrammatic representation of the unamputated Bethe-Salpeter amplitude for the pion.}
\label{fig:unam-BS}
\end{figure}

We start with the unamputated Bethe-Salpeter amplitude for the pion,
\begin{align}
    \chi_{\pi} (p;P)=G_{q}(p_{+})\,\Gamma_{\pi}(p;P)\,G_{q}(p_{-})\,, \label{eq:unam-BS}
\end{align}
with
\begin{align}
    p_{\pm}=p\pm P/2\,, \label{eq:ppm}
\end{align}
and the pion BS amplitude
\begin{align}
    \Gamma_{\pi}(p;P)=\mathrm{i}\gamma_{5}h_{\pi}(p;P)\,.
\end{align}
Here $p_{\pm}$ denotes the quark momenta and $P$ is the momentum of pion. The quark propagator is given by
\begin{align}
    G_{q} (p)=\big[-\mathrm{i}\gamma \cdot p +M_q(p)\big]\,\Delta\big(p^2,M_q(p)\big)\,,
\end{align}
with 
\begin{align}
    \Delta\big(p^2,M_q(p)\big)=\frac{1}{Z_{q}(p)}\frac{1}{p^{2}+M^{2}_{q}(p)}\,,\label{eq:quark-prop-scalar}
\end{align}
being a scalar function. In \labelcref{eq:quark-prop-scalar}, $Z_{q}(p)$ and $M_q(p)$ stand for the quark wave function and the quark mass function, respectively. The unamputated pion BS amplitude in \labelcref{eq:unam-BS} is also shown diagrammatically in \Cref{fig:unam-BS}.

%
%

The four-vector momentum $p$, which is both in the light-front and quasi-light-front formalisms, reads  
\begin{align}
    p_\mu=(p_{0},p_{3},p_{1},p_{2})=(p_{0},p_{3},p_{\perp})\,.\label{eq:pmu-def}
\end{align}
In \labelcref{eq:pmu-def}, the transverse momentum is given by $p_{\perp}=(p_{1},p_{2})$. As for the pion momentum, there is a slight difference. In the light-front formalism, the on-shell momentum of the pion in a rest frame is given by
\begin{align}
    P_{_{\mathrm{LF}},\mu}=(\mathrm{i} m_{\pi},0,0,0)\,,
\end{align}
with $P_{_{\mathrm{LF}}}^2=-m_{\pi}^2$. While in the quasi-light-front formalism, a nonzero longitudinal momentum for the pion, here denoted by $P_{z}$, is necessary. Thus, one arrives at 
\begin{align}
    P_\mu=(\mathrm{i}E_{\pi},P_{z},0,0)\,,\qquad E_{\pi}=\sqrt{P_{z}^{2}+m_{\pi}^{2}}\,, \label{eq:Pmu-Pz}
\end{align}
with the Lorentz invariant $P^2=-m_{\pi}^2$.

A twist-two quasi-parton distribution amplitude (quasi-PDA or qPDA) of pion is given by
\begin{align}
    &\phi_\pi(x, P_z)\nonumber\\[2ex]
    =&\frac{1}{f_\pi}\mathrm{Tr}_{\mathrm{CD}}\bigg[\int \frac{d^4 p}{(2\pi)^4}\delta(\tilde n\cdot p_+-x\tilde n\cdot P)\gamma_5 \gamma\cdot \tilde n \chi_\pi(p;P)\bigg]\,,\label{eq:qPDA}
\end{align}
with the pion decay constant $f_\pi$ and an Euclidean vector $\tilde n=(0,1,0,0)$, where the subscript $_{\mathrm{CD}}$ indicates that the trace operates in the color and Dirac space. The variable $x$ stands for the fraction of pion longitudinal momentum carried by the quark, that is implemented by the delta function. Note that if the vector $\tilde n$ is replaced with the lightlike vector $n=(\mathrm{i},1,0,0)$, the quasi-PDA in \labelcref{eq:qPDA} would return to the PDA on the light front. Moreover, the PDA can also be obtained from the quasi-PDA in the limit of large longitudinal momentum, i.e., $P_z \to \infty$, based on the LaMET \cite{Ji:2013dva, Ji:2020ect}, which is demonstrated and proved analytically in two simple cases in \Cref{app:pointlike,app:asymptotic}. If the transverse momentum in \labelcref{eq:qPDA} was not integrated and kept untouched, one would obtain instead the quasi-light-front wave function (quasi-LFWF) of pion, denoted by $\psi_\pi (x,P_{z},p_{\perp})$. Obviously, the quasi-PDA is related to the quasi-LFWF through the relation, as follows
\begin{align}
    \phi_\pi (x,P_{z})=\frac{1}{16 \pi^3}\int d^2 p_{\perp} \psi_\pi (x,P_{z},p_{\perp})\,.\label{eq:PDA-LFWF}
\end{align} 
In the same way, one can also obtain the valence-quark quasi-parton distribution function (quasi-PDF) for the pion \cite{Chang:2014lva, Chen:2016sno, Xu:2018eii}, as follows
\begin{align}
    u_{\pi}(x) =&\mathrm{Tr}_{\mathrm{CD}}\bigg[\int \frac{d^4 p}{(2\pi)^4}\delta(\tilde n\cdot p_+-x\tilde n\cdot P)\nonumber\\[2ex]
    &\times \Big (\tilde n\cdot \partial_p H_{+}(p;P)\Big) H_{-}(p;P)\bigg] - \mathcal{S}(x) \,,\label{eq:qPDF}
\end{align}
with
\begin{align}
    \mathcal{S}(x)=&\frac{1}{2}\mathrm{Tr}_{\mathrm{CD}}\bigg[\int \frac{d^4 p}{(2\pi)^4}\delta(\tilde n\cdot p_+-x\tilde n\cdot P)\nonumber\\[2ex]
    &\times \tilde n\cdot \partial_p \Big(H_{+}(p;P) H_{-}(p;P)\Big)\bigg]\,,\label{eq:qPDF2}
\end{align}
where one has $\tilde n\cdot \partial_p=\tilde n_\mu (\partial/\partial p_\mu)$ and
\begin{align}
    H_{+}(p;P)&=\bar \Gamma_{\pi}(p; -P) G_{q}(p_+)\,,\\[2ex]
    H_{-}(p;P)&=\Gamma_{\pi}(p; P) G_{q}(p_-)\,,
\end{align}
 with $\bar \Gamma_{\pi}(p; P)=C^\dagger \Gamma_{\pi}(p; P)^\mathrm{T}C$, where $C$ is the charge conjugation matrix.

\section{Quark propagator and Bethe-Salpeter amplitude}
\label{sec:fRG-input}

%
\begin{figure}[t]
\includegraphics[width=0.3\textwidth]{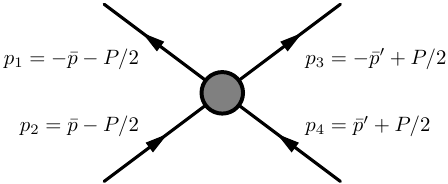}
\caption{Diagrammatic representation of the four-quark vertices, where all momenta are incoming.}
\label{fig:G4quark-Mom}
\end{figure}
%

Apparently, a pion BS amplitude as well as the quark propagator in the complex plane of momenta is required to compute the quasi-PDA in \labelcref{eq:qPDA} or the quasi-PDF in \labelcref{eq:qPDF} discussed in the section above. Recently, we have developed an approach of multi-quark flows within the functional renormalization group, that is well suited for the studies of bound state properties \cite{Fu:2022uow, Fu:2024ysj,Fu:2025hcm}. It is found that the flow of quark self-energy, i.e., the two-quark correlation function, and the flow of four-quark vertex play a similar role as the quark gap equation and Bethe-Salpeter equation for the four-quark scattering kernel, respectively. Note that the self-consistency between the two-quark and four-quark flows is guaranteed naturally by the fact that these two flows are derived from the same effective action, independent of truncations used.

In this work, we would like to calculate the pion quasi parton distributions based on the pion BS amplitude and quark propagator obtained from the infrared dynamics of four-quark scatterings. This low-energy effective theory (LEFT) within the fRG is introduced in \cite{Fu:2024ysj}, which can be directly embedded in the QCD, see \cite{Fu:2025hcm}. This also indicates that our approach can be directly applied to QCD. 

In this section, we briefly introduce the four-quark formalism and the related notations firstly, and then present the results of observables to be used. For more details, we refer to the works of series \cite{Fu:2022uow, Fu:2024ysj,Fu:2025hcm}. Let us begin with the effective action of a two-flavor LEFT, that is,
\begin{align}
    \Gamma_{k}[q, \bar{q}]=\Gamma_{\mathrm{kin},k}[q, \bar{q}]+\Gamma_{4q,k}[q, \bar{q}]\,,
\end{align}
being purely fermionic, where $k$ denotes the infrared cutoff, i.e., the renormalization group (RG) scale, in the flows of fRG. The kinetic term reads
\begin{align}
    \Gamma_{\mathrm{kin},k}=&\int \frac{d^4 p}{(2 \pi)^4} Z_{q}(p) \,\bar q(-p)\Big[\mathrm{i}\slashed{p} +M_q(p)\Big]\,q(p)\,.\label{eq:GammaKin}
\end{align}
with $\slashed{p}=\gamma_\mu p_\mu$. The four-quark term is given by
\begin{align}
    \Gamma_{4q,k} =&-\int\frac{d^4 p_1}{(2 \pi)^4}\cdots\frac{d^4 p_4}{(2 \pi)^4} (2\pi)^4\delta(p_1+p_2+p_3+p_4)\nonumber \\[2ex]
 & \times\lambda_{\alpha}(\boldsymbol{p}) \, {\mathcal{T}}^{(\alpha)}_{ijlm}(\boldsymbol{p})\,\bar q_i(p_1) q_j(p_2) \bar q_l(p_3) q_m(p_4)\,,\label{eq:Gamma4qGen}
\end{align}
where a Fierz-complete tensor basis of ten elements ${\mathcal{T}}^{(\alpha)}_{ijlm}$ for two-flavor quarks is used, and $\lambda_{\alpha}$ is the relevant four-quark dressing of tensor $\alpha$. A sum over the index $\alpha$ is assumed in \labelcref{eq:Gamma4qGen}.

%
\begin{figure}[t]
\includegraphics[width=0.48\textwidth]{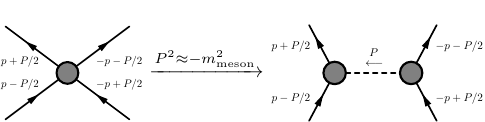}
\caption{Schematic demonstration of the decomposition of the four-quark vertex into the two quark-meson vertices connected by a meson propagator, when the $t$-channel momentum is near the pole mass of meson. The direction of external quark momenta is along with that of fermion flows.}
\label{fig:BSamplitude}
\end{figure}
%

The four-quark vertex is depicted in \Cref{fig:G4quark-Mom}, where the momentum notations are also shown. In our previous work \cite{Fu:2024ysj}, a three-momentum approximation for the four-quark vertices has been established, viz.,
\begin{align}
  \lambda_{\alpha}(p_1,p_2,p_3,p_4)\approx &\lambda_{\alpha}(s,t,u)\,, \label{eq:3chanAppro}
\end{align}
where the Mandelstam variables read
\begin{align}
    s=(\bar{p}+\bar{p}')^2,\quad t=P^2,\quad u=(\bar{p}-\bar{p}')^2\,.
\end{align}
The reliability of the approximation in \labelcref{eq:3chanAppro} was investigated in detail in \cite{Fu:2024ysj}, and it was found that the resulting error does not exceed 1.5\%. Being compatible with the three-momentum approximation, the momentum configuration of the four-quark vertex is chosen as 
\begin{align}
    P_\mu=&\sqrt{P^{2}}\,\Big(1,\,0,\,0,\,0 \Big)\,,\nonumber\\[2ex]
    \bar p_\mu=&\sqrt{p^{2}}\, \Big(1,\, 0,\,0,\,0 \Big)\,,\nonumber\\[2ex]
    \bar p^\prime_\mu=&\sqrt{p^{2}}\, \Big(\cos \theta,\, \sin \theta,\,0,\,0 \Big)\,,\label{eq:momframe1}
\end{align}
which are obviously in one-by-one correspondence to the Mandelstam variables $(s,t,u)$. When the $t$-channel momentum is on-shell, that is, the momentum is near the pole mass of mesons, e.g., $P^2 \approx -m_\pi^2$ for the pion channel, the strength of four-quark dressing increases remarkably, which indicates there is a resonance of emergent mesonic degrees of freedom. Consequently, in the resonance the four-quark vertex can be approximated by the two quark-meson vertices connected by a meson propagator as shown in \Cref{fig:BSamplitude}. In order to extract the BS amplitude appropriately, we choose another momentum configuration, as follows
\begin{align}
    P_\mu=&\sqrt{P^{2}}\,\Big(1,\,0,\,0,\,0 \Big)\,,\nonumber\\[2ex]
    \bar p_\mu=&\sqrt{p^{2}}\, \Big(\cos \theta,\, \sin \theta,\,0,\,0 \Big)\,,\nonumber\\[2ex]
    \bar p^\prime_\mu=&-\sqrt{p^{2}}\, \Big(\cos \theta,\, \sin \theta,\,0,\,0 \Big)\,.
    \label{eq:momframe2}
\end{align}
\Cref{eq:momframe2} is a bit different in comparison to \labelcref{eq:momframe1}. Here we have used $\bar p_\mu=-\bar p^\prime_\mu\equiv p_\mu$, which is in accordance with the definition of the BS amplitude in \Cref{fig:unam-BS}. 

Note that the computation above is done in the Euclidean regime, i.e., $t=P^2 \geq 0$. The Euclidean data of the four-quark dressings are continued analytically into the Minkowski regime based on the Pad\'e approximants, which are discussed in detail in \cite{Fu:2024ysj}. The pion pole mass is obtained from the zero point of the inverse of the four-quark coupling, viz.,
\begin{align}
    \lambda_\pi^{-1}(P^2=-m_\pi^2)=0\,.\label{eq:polemass}
\end{align}
The relevant BS amplitude just corresponds to the residue of the four-quark coupling at the pole, i.e.,
\begin{align}
    h_\pi(p,\cos \theta)=\lim_{P^2\to -m_\pi^2} \Big[\lambda_{\pi}(P^2,p,\cos \theta)\,(P^2+m_\pi^2)\Big]^{1/2}\,. \label{eq:BSa-pi}
\end{align}

As a short conclusion, we briefly introduced the LEFT with the fRG, which describes the infrared dynamics of four-quark scatterings, including its interaction truncations and momentum-dependent approximations. Based on this framework, we obtained the momentum-dependent quark mass function and the BS amplitude in Euclidean space. They will be applied in the following quasi-PDA and quasi-PDF calculations.

\section{Analytic structure in the complex plane}
\label{sec:Anal-struc} 

In this section, we analyze the complex structure of quasi-PDA and quasi-PDF and specifically introduce our approach. In \Cref{sec:integ-contour}, we introduce an integral contour with a finite shift in the imaginary direction by analyzing the distribution of quasi-PDA poles in the complex plane. In \Cref{sec:Taylor}, we introduce how to extract partial information in Minkowski space from the Euclidean quark mass function and BS amplitude through a Taylor expansion. This extends the computable range of $P_z$, leading to more precise PDA and PDF calculations, which are also shown in \Cref{sec:PDA} and \Cref{sec:PDF}.

\subsection{Deformed integration contour for the $p_0$-integral }
\label{sec:integ-contour} 

%
\begin{figure*}[t]
\includegraphics[width=0.45\linewidth]{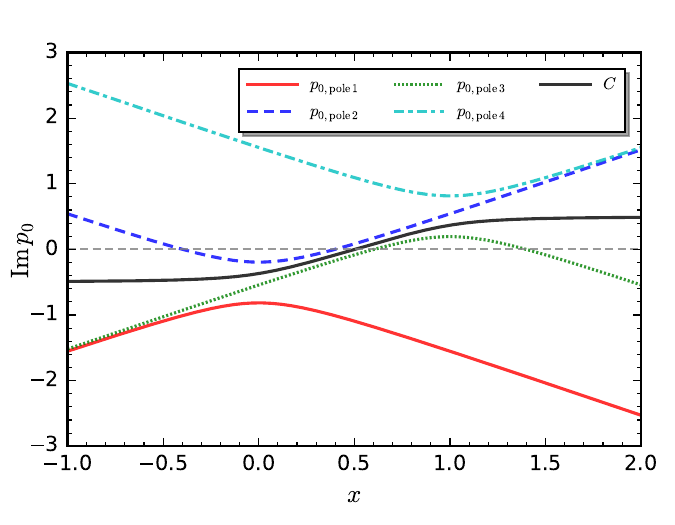}\hspace{0.3cm}
\includegraphics[width=0.45\linewidth]{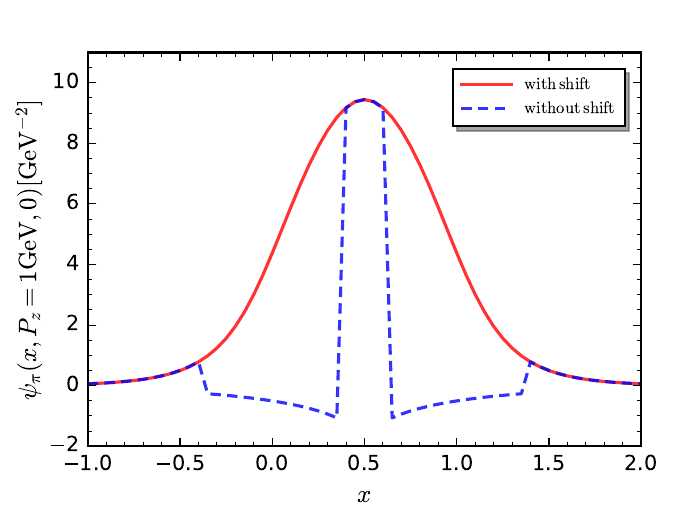}
\caption{Left panel: Imaginary parts of the four poles in \labelcref{eq:qPDA-algebraic} as functions of the momentum fraction $x$. The black solid line denotes the magnitude of the shift in the imaginary axis in the integral of $p_0$ for the quasi-PDA. Right panel: Quasi-light-front wave function of pion, $\psi_\pi$, as a function of the momentum fraction $x$. The result of the red solid line is obtained with a shift in the imaginary axis as shown in the left panel in the $p_0$-integral, while the blue dashed line corresponds to the result without such shift, i.e., integrating $p_0$ along the real axis. In both panels, the results are obtained with $p_{\perp}=0$, $P_{z}=1$ GeV, $M_q=0.35$ GeV, $m_\pi=0.14$ GeV, $h_\pi=1$.}
\label{fig:poles-LFWF}
\end{figure*}
%

Now it is ready to compute the quasi-PDA in \labelcref{eq:qPDA}. It is straightforward to integrate the longitudinal momentum $p_3$ by resorting to the delta function. Inserting \labelcref{eq:ppm,eq:Pmu-Pz}, one is immediately led to
\begin{align}
    p_3=\Big(x-\frac{1}{2}\Big)P_z\,, \label{eq:p3-Pz}
\end{align}
and the pion quasi-PDA
\begin{align}
    \phi_\pi(x, P_z)&=\frac{1}{f_\pi}\frac{4 N_c}{(2\pi)^4}\int d p_\perp^2 d p_0 \, h_\pi(p,\cos \theta) P_z \nonumber\\[2ex]
    &\times \frac{1}{Z_q(p_+)Z_q(p_-)}\frac{1}{p_+^2+M_q^2(p_+)}\frac{1}{p_-^2+M_q^2(p_-)}\nonumber\\[2ex]
    &\times \Big [x M_q(p_-)+(1-x)M_q(p_+)\Big] \,,\label{eq:qPDA-algebraic}
\end{align}
with
\begin{align}
    \cos \theta=\frac{p_0}{p}, \qquad p=\sqrt{p_0^2+p_3^2+p_\perp^2}\,. \label{}
\end{align}

Before we integrate the variable $p_0$, it is necessary to analyze the analytic structure of \Cref{eq:qPDA-algebraic}. To that end, in this section, we assume for the moment that both the BS amplitude $h_\pi$ and the quark mass $M_q$ are constants, which allows us to do analytic calculations. Obviously from the propagators of the quark and anti-quark in \labelcref{eq:qPDA-algebraic}, one readily finds four poles on the purely imaginary axis. For the quark, the two poles are labeled as
\begin{subequations} 
\label{eq:poles}
\begin{align}
     p_{0,\mathrm{pole}\,1}=&\mathrm{i}\bigg[-\frac{1}{2}\sqrt{P_z^2+m_\pi^2}-\sqrt{p_\perp^2+(xP_z)^2+M_q^2}\bigg]\,,\nonumber\\[2ex]
     p_{0,\mathrm{pole}\,2}=&\mathrm{i}\bigg[-\frac{1}{2}\sqrt{P_z^2+m_\pi^2}+\sqrt{p_\perp^2+(xP_z)^2+M_q^2}\bigg]\,.\label{eq:quark-poles}
\end{align}
The other two poles for the anti-quark read
\begin{align}
     p_{0,\mathrm{pole}\,3}=&\mathrm{i}\bigg[\frac{\sqrt{P_z^2+m_\pi^2}}{2}-\sqrt{p_\perp^2+\big((1-x)P_z \big)^2+M_q^2}\bigg]\,,\nonumber\\[2ex]
     p_{0,\mathrm{pole}\,4}=&\mathrm{i}\bigg[\frac{\sqrt{P_z^2+m_\pi^2}}{2}+\sqrt{p_\perp^2+\big((1-x)P_z \big)^2+M_q^2}\bigg]\,.\label{eq:antiquark-poles}
\end{align}
\end{subequations}
It can be easily found that the first pole is always in the lower half complex plane and the fourth pole always in the upper half. What is intriguing is that the second pole, supposed to be in the upper half plane, might cross over the $x$-axis and enter into the lower half if the transverse momentum $p_\perp$ is small while $P_z$ is large. In the same way, the third pole
could move from the lower half to the upper half plane. This behavior is clearly shown in the left panel of \Cref{fig:poles-LFWF}, where the imaginary parts of the four poles in \labelcref{eq:poles} are depicted as functions of the momentum fraction $x$. Here, we have chosen $p_{\perp}=0$, $P_{z}=1$ GeV, $M_q=0.35$ GeV, $m_\pi=0.14$ GeV for illustrative purpose. One can see that the second pole enters into the lower half plane in the regime of $-0.5< x< 0.5$ and the third pole into the upper half plane in the regime of $0.5< x< 1.5$.

Note that when the crossing takes place, the $p_0$-integral, being originally along the real axis, should be modified accordingly. Otherwise the contribution of quark or anti-quark to \Cref{eq:qPDA-algebraic} could not be taken into account. This is demonstrated clearly in the right panel of \Cref{fig:poles-LFWF}, where the quasi-LFWF of pion on the right side of \labelcref{eq:PDA-LFWF} is presented. The blue dashed line denotes the results with the $p_0$-integral along the real axis. In the two disconnected crossing regions with $-0.5< x< 0.5$ and $0.5< x< 1.5$, being symmetric with respect to $x=0.5$, the integrated results are obviously incorrect. To correct the results, one just needs to shift the integral to that with a finite imaginary part, i.e.,
\begin{align}
    \int_{-\infty}^{\infty}d p_0 \to \int_{-\infty +\mathrm{i} C}^{\infty +\mathrm{i} C}d p_0\,, \label{eq:integral-shift}
\end{align}
such that the shifted integral contour crosses between the first and second poles in \labelcref{eq:quark-poles} and between the third and fourth poles in \labelcref{eq:antiquark-poles}. This requires
\begin{align}
    \mathrm{Im}\, p_{0,\mathrm{pole}\,1}&<C<\mathrm{Im}\, p_{0,\mathrm{pole}\,2},\nonumber\\[2ex]
    \mathrm{Im}\, p_{0,\mathrm{pole}\,3}&<C<\mathrm{Im}\, p_{0,\mathrm{pole}\,4}\,, \label{eq:deform-cont}
\end{align}
where $C$ is the shifted magnitude in \labelcref{eq:integral-shift}. Here we choose 
\begin{align}
    C=\frac{\mathrm{Im}\,p_{0,\mathrm{pole}\,2}+\mathrm{Im}\,p_{0,\mathrm{pole}\,3}}{2}\,,\label{eq:shifted-magn}
\end{align}
as shown by the black solid line in the left panel of \Cref{fig:poles-LFWF}, which obviously satisfies the constraints in \labelcref{eq:deform-cont}. The resulting quasi-LFWF after the shift is given by the red solid line in the right panel of \Cref{fig:poles-LFWF}, which are the desired correct results. Note that the results of the red solid and blue dashed lines are identical to each other when there is no crossing as they should be.

\subsection{Taylor expansion of the Bethe-Salpeter amplitude and quark mass in the complex plane}
\label{sec:Taylor}

As we have discussed in \Cref{sec:integ-contour}, the shift 
\begin{align}
    p_0 \to p_0 + \mathrm{i} C\,,
\end{align}
is required in the $p_0$-integral, which allows us to take into account the contributions from the poles of quark and anti-quark properly. This entails that the momentum $p$ in the BS amplitude in \labelcref{eq:BSa-pi} has the form as follows
\begin{align}
    p_\mu=\Big(p_{0}+ \mathrm{i} C,(x-1/2)P_z,p_{\perp}\Big)\,,\label{eq:p-complex}
\end{align}
where we have used the relation in \labelcref{eq:p3-Pz} arising from the delta function in \labelcref{eq:qPDA}. In the same way, the momenta of quark and anti-quark as shown in \Cref{fig:unam-BS} or \labelcref{eq:qPDA-algebraic} read
\begin{align}
     p_{+\mu}=&\Big(p_{0}+ \mathrm{i}  \big(C+\sqrt{P_z^2+m_\pi^2}/2\big),x P_z,p_{\perp}\Big)\,,\nonumber\\[2ex]
     p_{-\mu}=&\Big(p_{0}+\mathrm{i} \big(C-\sqrt{P_z^2+m_\pi^2}/2 \big),(x-1) P_z,p_{\perp}\Big)\,.\label{eq:pm-complex}
\end{align}
Evidently, the zeroth components in all the momenta from \labelcref{eq:p-complex,eq:pm-complex} are complex-valued. In this work, we use the Taylor expansion to extend the momentum dependence to the complex plane, by means of a series of powers of the imaginary part. This leads us to
\begin{align}
    h_\pi(p,\cos \theta)=&h_\pi(\bar p,\cos \theta)+\sum_{m=1}^{n}\frac{1}{m !}\frac{\partial^m h_\pi}{\partial p_0^m} (\Delta p_0)^m\nonumber\\[2ex]
    &+\mathcal{O} \big((\Delta p_0)^{m+1}\big)\,,\label{eq:h-expan}
\end{align}
with 
\begin{align}
    \bar p_\mu=\Big(p_{0},(x-1/2)P_z,p_{\perp}\Big)\,,\quad \Delta p_0=\mathrm{i} C\,,\label{eq:barp}
\end{align}
for the pion BS amplitude, where it is expanded to the $n$-th order. Similarly one finds the quark mass
\begin{align}
    M_q(p_+)=&M_q(\bar p_+)+\sum_{m=1}^{n}\frac{1}{m !}\frac{\partial^m M_q}{\partial p_0^m} (\Delta p^{+}_{0})^m\nonumber\\[2ex]
    &+\mathcal{O} \big((\Delta p^{+}_{0})^{m+1}\big)\,,\nonumber\\[2ex]
    M_q(p_-)=&M_q(\bar p_-)+\sum_{m=1}^{n}\frac{1}{m !}\frac{\partial^m M_q}{\partial p_0^m} (\Delta p^{-}_{0})^m\nonumber\\[2ex]
    &+\mathcal{O} \big((\Delta p^{-}_{0})^{m+1}\big)\,,\label{eq:Mq-expan}
\end{align}
with
\begin{align}
     \bar p_{+\mu}=\Big(p_{0},x P_z,p_{\perp}\Big)\,,\quad
     \bar p_{-\mu}=\Big(p_{0},(x-1) P_z,p_{\perp}\Big)\,,\label{eq:barppm}
\end{align}
and 
\begin{align}
     \Delta p^{+}_{0}=&\mathrm{i}  \big(C+\sqrt{P_z^2+m_\pi^2}/2\big)\,,\nonumber\\[2ex]
     \Delta p^{-}_{0}=&\mathrm{i} \big(C-\sqrt{P_z^2+m_\pi^2}/2 \big)\,.\label{}
\end{align}
In the numerical calculations in this work, the Taylor expansion for the BS amplitude and the quark masses in \labelcref{eq:h-expan,eq:Mq-expan} has been done up to the fourth order.

\section{Numerical results}
\label{sec:numerical}

%
\begin{figure}[t]
\includegraphics[width=0.45\textwidth]{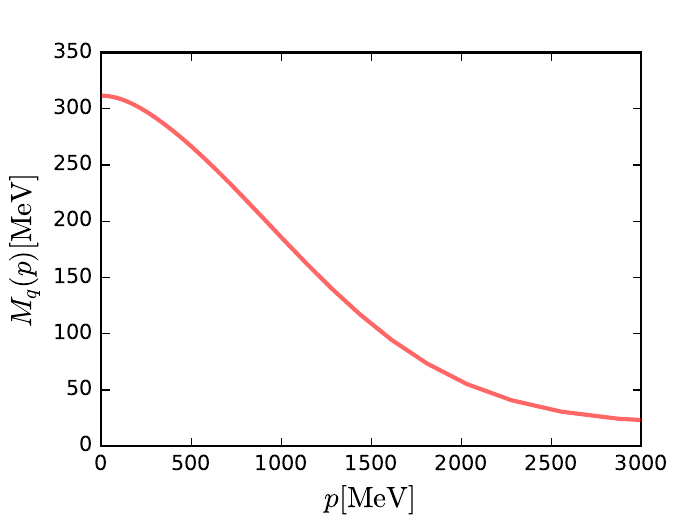}
\caption{Light constituent quark mass $M_q(p)$ as a function of the momentum obtained with the two-flavor low energy effective theory within the functional renormalization group approach in \cite{Fu:2024ysj}.} 
\label{fig:Mq}
\end{figure}
%

%
\begin{figure}[t]
\includegraphics[width=0.48\textwidth]{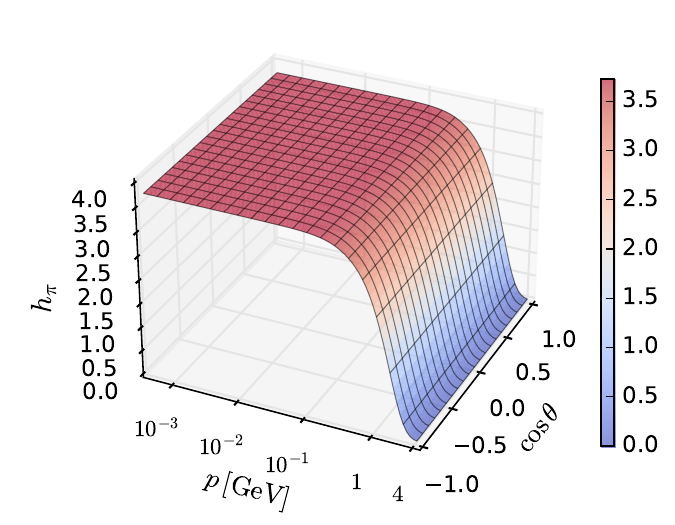}
\caption{3D plot of the Bethe-Salpeter amplitude of pions as a function of the magnitude of quark momentum and the angle between the quark and pion momenta, as shown in \labelcref{eq:BSa-pi}. The results are obtained with the two-flavor low energy effective theory within the functional renormalization group approach in \cite{Fu:2024ysj}.}\label{fig:hpi}
\end{figure}
%

\subsection{Quark mass and Bethe-Salpeter amplitude}
\label{sec:Mq-hpi}

A four-quark low energy effective theory within the fRG was established in \cite{Fu:2024ysj}. In this approach quantum fluctuations of different momentum modes are integrated successively with the evolution of RG flows \cite{Fu:2022gou, Dupuis:2020fhh}. The flows are initiated from an ultraviolet (UV) cutoff scale, denoted by $k=\Lambda$, and evolved towards the infrared (IR). A full quantum effective action is resolved after the IR limit $k=0$ is arrived at \cite{Wetterich:1992yh}. In the LEFT at the initial scale $\Lambda$ the four-quark couplings of the pion and sigma channels are given by 
\begin{align}
    \lambda_{\sigma,\Lambda}(\bar p)=\lambda_{\pi,\Lambda}(\bar p)=\bar{\lambda}\,\Lambda^{-2} \exp\left[-\frac{1}{2}\left(\frac{\bar{p}}{ \Lambda}\right)^2 \right]\,,\label{eq:ini-lamsp}
\end{align}
with the average momentum
\begin{align}
    \bar{p}=\sqrt{p_{1}^2+p_{2}^2+p_{3}^2+p_{4}^2}\,,
\end{align}
while the four-quark couplings of other channels among the ten Fierz-complete tensor structures are vanishing at $k=\Lambda$. The UV cutoff $\Lambda$, $\bar{\lambda}$ in \labelcref{eq:ini-lamsp} and the light quark mass $M_{q,\Lambda}$ at $k=\Lambda$ constitute the parameters of the two-flavor LEFT. They are fixed by fitting the pion mass $m_{\pi}=138$ MeV and the pion decay constant $f_{\pi}=93$ MeV, and their values read
\begin{align}
    \Lambda=850\,\mathrm{MeV},\quad M_{q,\Lambda}=19.5\,\mathrm{MeV},\quad\bar{\lambda}=16.7\,.\label{eq:initial-condition}
\end{align}
For more details about the LEFT and its parameters, we refer to \cite{Fu:2024ysj}.

In \Cref{fig:Mq} we show the quark mass function obtained in the two-flavor LEFT within the fRG. This result is also compared with the lattice results in \cite{Fu:2024ysj}. It was found there that the quark mass function obtained in the LEFT is qualitatively consistent with the lattice quark mass in the Landau gauge. But there was still some quantitative difference: More specifically, the dependence of the quark mass in LEFT is relatively flatter in comparison to the lattice QCD. This is attributed to the fact that the gluon dynamics, that are missing in the LEFT, have already begun to play a role in the momentum regime around the UV cutoff of LEFT $\Lambda=850$ MeV. This was also confirmed in \cite{Fu:2024ysj} with the observation that when the UV cutoff is reduced, say $\Lambda=500$ MeV, better agreement for the quark mass is found. Furthermore, it was found that the quark wave function $Z_q(p)$ in \labelcref{eq:quark-prop-scalar} was very close to unity, and the difference $Z_q(p)-1$ was negligible in the LEFT. See \cite{Fu:2024ysj} for more details.

The pion BS amplitude calculated in \cite{Fu:2024ysj} is also presented in \Cref{fig:hpi}. One can see that the dependence on the angle between the quark and meson momenta is very mild in the LEFT, in contradistinction to the case of QCD where a nontrivial dependence of the BS amplitude on the angle is found.

\subsection{Quasi-LFWF, quasi-PDA, PDA and its moments}
\label{sec:PDA}

%
\begin{figure}[t]
\includegraphics[width=0.45\textwidth]{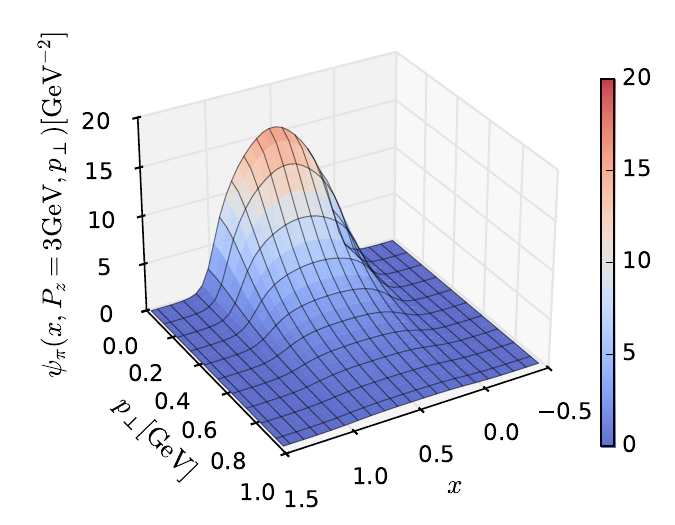}
\caption{3D plot of the pion quasi-LFWF as a function of the momentum fraction $x$ and transverse momentum $p_{\perp}$ with the longitudinal momentum $P_z=3$ GeV.}\label{fig:quasi-LFWF}
\end{figure}
%

%
\begin{figure}[t]
\includegraphics[width=0.45\textwidth]{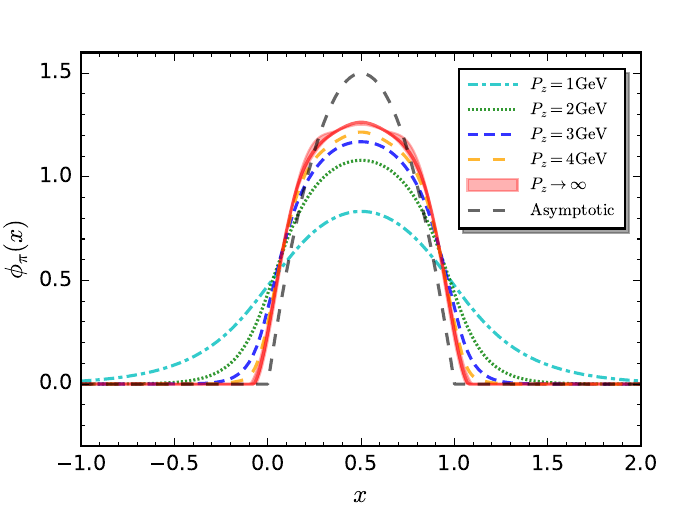}
\caption{Quasi-PDA of pions as a function of the momentum fraction $x$ calculated with several different finite values of $P_z$. The red solid line with error bands denotes the extrapolated PDA with $P_z \to \infty$ from \labelcref{eq:PDA-extrap} based on the LaMET. The error arises from the region choice of the longitudinal momentum $P_{z}$ for the extrapolation, cf., the text for the details. The asymptotic PDA labeled by the gray dashed line is also presented for comparison.}\label{fig:quasi-PDA}
\end{figure}
%

%
\begin{figure}[t]
\includegraphics[width=0.45\textwidth]{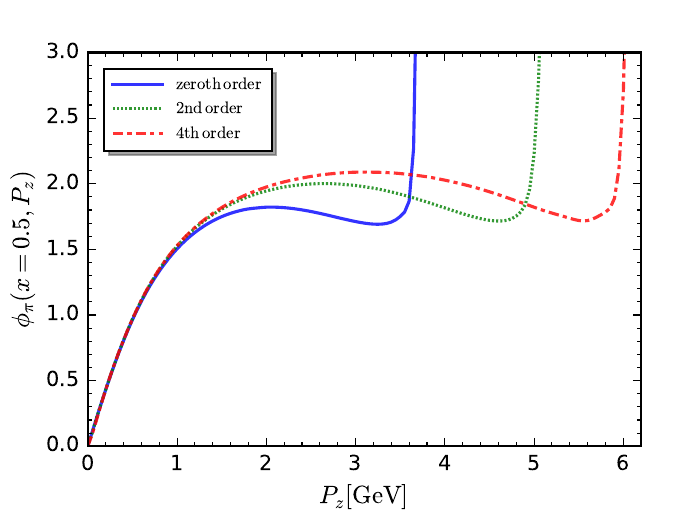}
\caption{Non-normalized quasi-PDA $\phi_{\pi}$ as a function of the longitudinal momentum $P_{z}$ at $x=0.5$. Results shown in different colors are obtained from the quark mass function and BS amplitude at different orders of the Taylor expansion, see \labelcref{eq:h-expan,eq:Mq-expan}.}\label{fig:pda-pz}
\end{figure}
%

In this subsection, we present the calculated results of quasi-parton distribution amplitude for the pions. But before we do that, it is desirable to show its quasi-light-front wave function, from which the quasi-PDA is immediately obtained by integrating out the dependence of transverse momentum, cf. \labelcref{eq:PDA-LFWF}. The results of quasi-LFWF are presented in \Cref{fig:quasi-LFWF}, where the longitudinal momentum of pion $P_z=3$ GeV is chosen. It can be found that the wave function $\psi_\pi$ is symmetric with respect to the plane $x=0.5$, and the height of the bump decreases with the increasing transverse momentum.

In \Cref{fig:quasi-PDA} we show the pion quasi-PDA as a function of $x$ with values of the longitudinal momentum of pion $P_z$ chosen from 1 to 4 GeV. Certainly, the larger the longitudinal momentum $P_z$, the more advantageous it is to use the quasi-PDA to extract the light-front PDA based on the LaMET. Nonetheless, there is usually a maximal value of $P_z$ in calculations, and only below this maximal longitudinal momentum the computation is applicable. The underlying reason is illustrated as follows. 

We begin with the comparison of relative positions of the second and third poles in the complex plane as shown in \labelcref{eq:quark-poles,eq:antiquark-poles}. If we assume that the imaginary part of the second pole is larger than that of the third pole, one readily finds from \Cref{eq:quark-poles,eq:antiquark-poles}
\begin{align}
    4 p_\perp^2+(2M_q)^2 >m_\pi^2\,,\label{eq:bound-constr}
\end{align}
with $x=1/2$. This constrain can be further reduced to
\begin{align}
    2M_q >m_\pi\,,\label{eq:bound-constr2}
\end{align}
with $p_\perp=0$. \Cref{eq:bound-constr2} is always satisfied if the quark mass is a constant and not dependent on the momentum, as shown in the left panel of \Cref{fig:poles-LFWF}. Note that the poles in \labelcref{eq:poles} are also obtained with the assumption of constant quark mass. The constraint in \labelcref{eq:bound-constr2} in fact is the stability condition for the pion, otherwise the pion would dissolve into a pair of quark and anti-quark since the binding energy is positive from the naive viewpoint. However, If we include the momentum dependence of the quark mass further, as shown in \labelcref{eq:Mq-expan}, the constraint in \labelcref{eq:bound-constr2} could be violated with the increase of $P_z$. This is confirmed in our numerical calculations.

In \Cref{fig:pda-pz} we show the dependence of the quasi-PDA of pion on the longitudinal momentum $P_z$ with $x=0.5$ fixed, where results obtained from different orders of Taylor expansion are presented. Note that the quasi-PDA here is not normalized, that is, its zeroth cumulant in \labelcref{eq:cumulants} is not necessarily unity. One can see that $\phi_\pi$ increases with $P_z$ in the region of small longitudinal momentum, and then is flattened and a plateau is developed, which indicates that the saturation of quasi-PDA with the increasing $P_z$ is approached. The saturation momentum where the plateau appears increases from $\sim 2$ GeV to $\sim 4$ GeV, when the order of Taylor expansion is increased from the zeroth to the fourth. The $\phi_\pi$ drops slightly after the $P_z$ exceeds the saturation momentum and then increases abruptly due to the reason discussed above, which indicates that the longitudinal momentum has already been located out of the region of reliability. Note that the slight drop above the saturation momentum is due to the unrealistic large quark mass in the region of ultraviolet large momentum in the LEFT, cf. \Cref{fig:Mq}, which would disappear in QCD. The abrupt divergence of the $\phi_\pi$ at large momentum stems from the restriction of the reliable calculation range for the longitudinal momentum, see \labelcref{eq:bound-constr} and \labelcref{eq:bound-constr2}. The improvement in the reliable range requires incorporating more information about the quark mass function and BS amplitude in the complex plane.

%
\begin{figure}[t]
\includegraphics[width=0.45\textwidth]{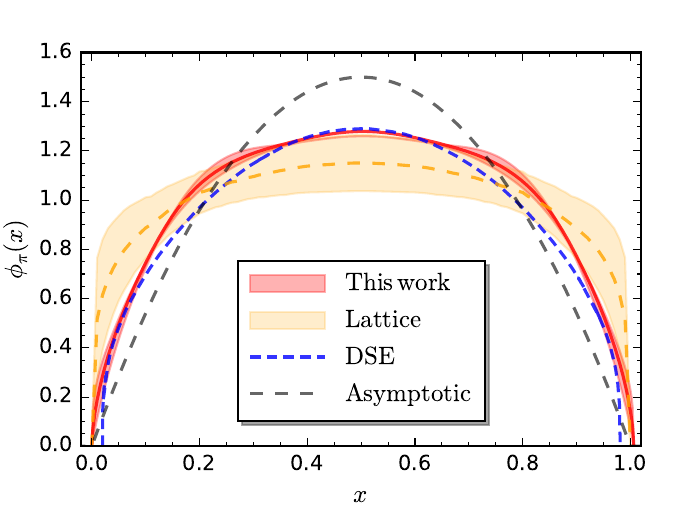}
\caption{Light-front PDA of pion as a function of the momentum fraction $x$. Our results are compared with those of lattice QCD \cite{LatticeParton:2022zqc} and Dyson-Schwinger equations \cite{Chang:2013pq}. The asymptotic result $6x(1-x)$ is also presented. The error estimate of our results is indicated by the band around the red solid line.}\label{fig:PDA}
\end{figure}
%

%
\begin{table*}[t]
  \begin{center}
  \begin{tabular}{lccc}
    \hline\hline & & &   \\[-2ex]   
    Method & $\langle\xi^2\rangle_\pi$ & $\langle\xi^4\rangle_\pi$ & $\langle\xi^6\rangle_\pi$   \\[1ex]
    \hline & & &   \\[-2ex]
    fRG (This Work) & $0.244^{+4}_{-6}$ & $0.119^{+5}_{-7}$ & $0.072^{+5}_{-6}$   \\[1ex]
    Lattice LaMET (LPC) \cite{LatticeParton:2022zqc} & $0.300(41)$ & - & -   \\[1ex]
    DSE \cite{Chang:2013pq, Roberts:2021nhw} & $0.251$ & $0.128$ & -   \\[1ex]
    Lattice OPE (RBC and UKQCD) \cite{Arthur:2010xf} & $0.28(1)(2)$ & - & -   \\[1ex]
    Lattice OPE (RQCD) \cite{RQCD:2019osh} & $0.234^{+6}_{-6}(4)(4)(2)$ & - & -   \\[1ex]
    Sum Rule \cite{Ball:2007zt, Zhong:2021epq}& $0.271(13)$ & $0.138(10)$ & $0.087(6)$   \\[1ex]
    \hline\hline
  \end{tabular}
  \caption{Cumulants of the valence-quark parton distributions of the pion up to the sixth order calculated in this work. The results are compared with those of lattice QCD based on the LaMET \cite{LatticeParton:2022zqc} and the DSE \cite{Chang:2013pq, Roberts:2021nhw}. Moreover, we also present the results of lattice QCD based on the Operator Product Expansion (OPE) \cite{Arthur:2010xf, RQCD:2019osh} and QCD sum rules \cite{Ball:2007zt, Zhong:2021epq}.}
  \label{tab:cumulants}
  \end{center}\vspace{-0.5cm}
\end{table*}
%

Due to the reasons discussed above, the longitudinal momentum of the quasi-PDA in our calculations is restricted to the region of $P_z\lesssim 4$ GeV and the Taylor expansion is done up to the fourth order. Evidently, it is found in \Cref{fig:quasi-PDA} that one has already arrived at saturation when $P_z$ is increased up to 4 GeV. Moreover, in the same plot we also present the results of light-front PDA extracted from the calculated quasi-PDA based on the LaMET \cite{Ji:2013dva, Ji:2020ect}, that is
\begin{align}
     \phi_\pi(x, P_z)=\phi_\pi(x, P_z\to \infty)+\frac{c_2(x)}{P_z^2}+\frac{c_4(x)}{P_z^4}+\mathcal{O}\Big(\frac{1}{P_z^6}\Big)\,,\label{eq:PDA-extrap}
\end{align}
where it is expanded to the fourth order of $1/P_z$. Here $\phi_\pi(x, P_z\to \infty)$ corresponds to the PDA of pion, which is denoted specifically by
\begin{align}
     \tilde \varphi_\pi(x)=\phi_\pi(x, P_z\to \infty)\,.\label{eq:PDA-m-pi}
\end{align}
The error band of PDA is also shown in \Cref{fig:quasi-PDA}, which estimates the errors of the extracted PDA arising from different momentum regions used to do the extrapolation, $P_z\in [1~\mathrm{GeV}, P_z^{\mathrm{max}}]$ with $P_z^{\mathrm{max}}=3, 4, 5$ GeV.

Note that in the endpoint regions of momentum fraction, i.e., in the proximity of $x\sim 0$ or  $x\sim 1$, LaMET cannot be used reliably \cite{Ji:2020ect}, which is reflected by the non-physical results of $\tilde \varphi_\pi(x)$ \labelcref{eq:PDA-m-pi} found in \Cref{fig:quasi-PDA}, where $\tilde \varphi_\pi(x)$ is nonzero when the fraction $x$ is a bit smaller than zero or larger than unity, although it indeed decays into zero rapidly as one approaches more deeply towards the outer regions. In order to circumvent this problem, we adopt a phenomenological method as used in \cite{LatticeParton:2022zqc}, where a further extrapolation 
\begin{align}
     \tilde \varphi_\pi(x)\xrightarrow{c x^a (1-x)^b}\varphi_\pi(x)\,,\label{eq:PDA-pi}
\end{align}
in the endpoint regions with the fitting parameters $a,\,b,\,c$ is implemented. The endpoint regions are chosen as
\begin{align}
     0 \leq x \leq x_{\mathrm{EP}}\,,\quad\mathrm{and}\quad 1-x_{\mathrm{EP}} \leq x \leq 1\,,\label{eq:endpoint-reg}
\end{align}
with $x_{\mathrm{EP}}=0.1$. We also vary the range of endpoint regions, e.g., $x_{\mathrm{EP}}=0.05, 0.15$ to estimate the errors of the extrapolation in \labelcref{eq:PDA-pi}. The final result of $\varphi_\pi(x)$ is presented in \Cref{fig:PDA} by the red solid line with its error band. The errors include both the error estimate in the extrapolation of $P_z$ in \labelcref{eq:PDA-extrap} and that of $x$ in \labelcref{eq:PDA-pi}. In \Cref{fig:PDA} we also compare our results with the computations of lattice QCD \cite{LatticeParton:2022zqc} and Dyson-Schwinger equations (DSE) \cite{Chang:2013pq}. It is found that our calculated pion PDA is in qualitative agreement with the lattice result within errors. Moreover, one can see that the pion PDA as a function of the momentum fraction calculated in this work is thinner than the central value of lattice QCD, and comparable with the result of DSE.

The cumulants of PDA are very useful in quantifying the parton distributions, which are defined as
 \begin{align}
    \langle\xi^n\rangle_\pi \equiv &\int_0^1 dx \, \xi^n \varphi_\pi(x)\,,\quad \mathrm{with} \quad \xi=2x-1\,, \label{eq:cumulants}
\end{align}
where $n$ denotes the order of cumulants. Since the pion PDA is symmetric with respect to $x=1/2$, as shown in \Cref{fig:PDA}, the odd-order cumulants of pion are vanishing. Our calculated cumulants of pion PDA up to the sixth order are shown in \Cref{tab:cumulants}, which are obtained from the PDA in \Cref{fig:PDA}. Moreover, several other calculations from lattice QCD, DSE, QCD sum rules, etc., are also presented in \Cref{tab:cumulants} for comparison. It is found that the quadratic cumulant of fRG is comparable with the DSE result \cite{Chang:2013pq, Roberts:2021nhw} and the lattice based on the Operator Product Expansion (OPE) \cite{RQCD:2019osh}, but slightly smaller than the lattice with the LaMET \cite{LatticeParton:2022zqc}. The higher-order cumulants obtained in this work, $\langle\xi^4\rangle_\pi$ and $\langle\xi^6\rangle_\pi$, are consistent with the DSE and QCD sum rules \cite{Ball:2007zt, Zhong:2021epq}.

\subsection{quasi-PDF and PDF}
\label{sec:PDF}

%
\begin{figure}[t]
\includegraphics[width=0.45\textwidth]{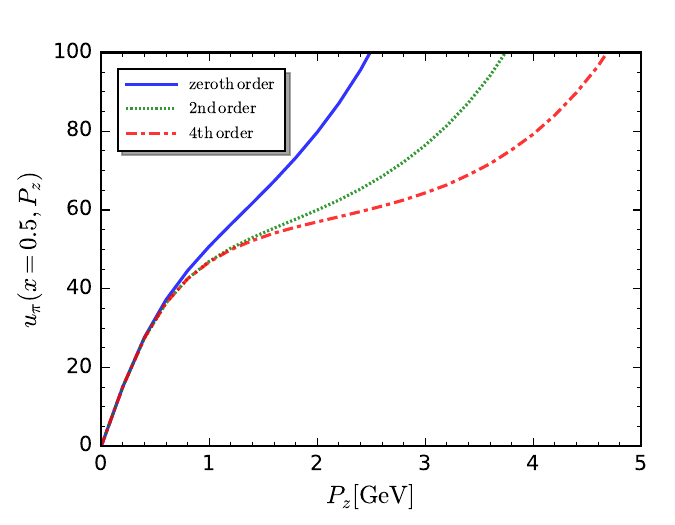}
\caption{Non-normalized valence $u$-quark quasi-PDF of pion as a function of the longitudinal momentum $P_{z}$ at $x=0.5$. Results obtained in different orders of Taylor expansion for the quark mass function and BS amplitude are shown in different colors.}\label{fig:pdf-pz}
\end{figure}
%

%
\begin{figure}[t]
\includegraphics[width=0.45\textwidth]{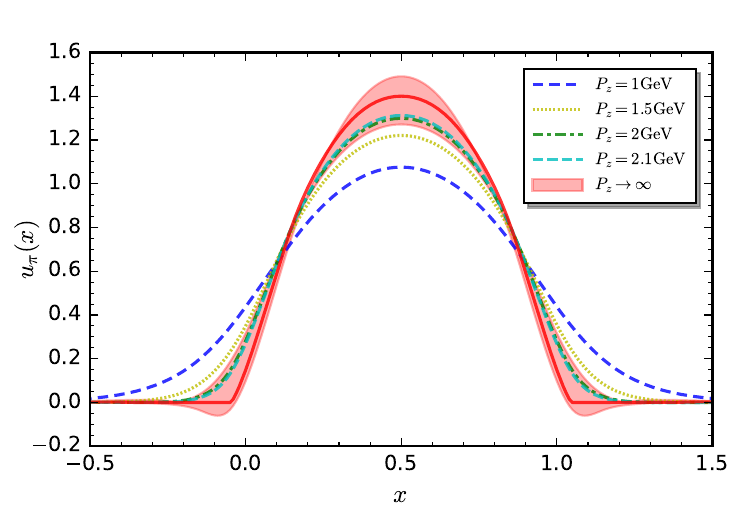}
\caption{Valence $u$-quark quasi-PDF of pion as a function of the momentum fraction $x$ calculated with several different finite values of $P_z$. The red solid line with error bands denotes the extrapolated PDF with $P_z \to \infty$ based on the LaMET.}\label{fig:PDF}
\end{figure}
%

In this subsection we present and discuss the results of the valence-quark quasi-PDF of pion calculated from \labelcref{eq:qPDF}. First of all, similar to \Cref{fig:pda-pz} for the quasi-PDA in \Cref{fig:pdf-pz} we show the dependence of quasi-PDF on $P_{z}$ at $x=0.5$. In the same way, results of Taylor expansion in \labelcref{eq:h-expan,eq:Mq-expan} up to different orders are compared. Comparing \Cref{fig:pdf-pz} with \Cref{fig:pda-pz}, one can find that it is more difficult to achieve a saturation momentum in quasi-PDF than quasi-PDA. This can be understood readily since the momentum derivative in \labelcref{eq:qPDF,eq:qPDF2} increases the power of the quark propagator in quasi-PDF in comparison to the quasi-PDA in \labelcref{eq:qPDA}, which would make the influence in \labelcref{eq:bound-constr,eq:bound-constr2} more sizable. Consequently, we choose the saturation momentum in \Cref{fig:pdf-pz} as the $P_z$ which corresponds to the minimal slope. For instance, the saturation momentum is $\sim 2.1$ GeV in the case of Taylor expansion up to the quartic order, which is significantly smaller than the counterpart $\sim 4$ GeV for the quasi-PDA in \Cref{fig:pda-pz}.

The valence $u$-quark quasi-PDF of pion at several values of $P_z$ is depicted as a function of $x$ in \Cref{fig:PDF}. The  PDF of pion on the light front is extrapolated from these quasi-PDFs through a similar LaMET relation in \labelcref{eq:PDA-extrap} with $P_z\to \infty$. As same as the extrapolation of PDA in \Cref{fig:quasi-PDA}, we vary the momentum range to estimate the errors of extrapolation, which is shown in the error band in \Cref{fig:PDF}. Evidently, the errors of PDF are significantly larger than those of PDA, since the region of feasible momentum $P_z$ for the quasi-PDF shrinks remarkably in contrast to that of quasi-PDA, as discussed above. This unreliability comes from the divergence of higher-order poles on the right-hand side of \labelcref{eq:qPDF}, which requires us to consider the quark mass function and BS amplitude in the complex plane more precisely.

\section{Conclusions and outlook}
\label{sec:conclusion}

In this work, we have developed a method within the functional renormalization group approach to calculate the valence quark quasi-PDA and quasi-PDF for the pion with a large longitudinal momentum. This method is applied in a low-energy effective theory with four-quark scatterings. The calculated quasi-PDA or quasi-PDF is used to extrapolate the light-front PDA or PDF based on the large momentum effective theory (LaMET).

The flows of the quark propagator and the four-quark vertices constitute a closed set of self-consistent equations, which can be regarded as analogues of the gap equation for the quark propagator and the Bethe-Salpeter equation for the four-quark scattering. Based on this observation and related advantages, we have constructed a framework of multi-quark flows within the fRG to study properties of bound states \cite{Fu:2022uow, Fu:2024ysj,Fu:2025hcm}. More specifically, taking the pion for example, one is able to obtain its pole mass and Bethe-Salpeter amplitude by resolving the pole of the four-quark vertex of pion channel and the residue at the pole.

When one adopts the LaMET method to calculate the light-front PDA or PDF by extrapolating the data of quasi-PDA or quasi-PDF from finite $P_z$ to $P_z \to \infty$, where $P_z$ is the longitudinal momentum of pion, there is a key point, that is, can one push $P_z$ to a value as large as possible? However, in realistic calculations the maximal $P_z$, below which the computation of quasi-PDA or quasi-PDF is applicable, is usually constrained by our knowledge of ingredients of parton distribution, e.g., the quark propagator and the pion BS amplitude, on the complex plane. In order to incorporate analytic information of parton distributions as much as possible, we have employed two methods in this work. One is the deformed integration contour that allows us to encode the pole contribution of quark propagators correctly. The other one is the Taylor expansion for the BS amplitude and quark mass.

We find that the pion light-front PDA extracted from the quasi-PDA in the low energy effective theory in this work is in qualitative agreement with the lattice QCD within errors. Moreover, the quadratic cumulant of pion PDA obtained in this work is comparable with lattice QCD and Dyson-Schwinger equation.

As discussed in \Cref{sec:fRG-input}, the approach introduced in this work is also applicable to the inputs of other momentum-dependent quark two-point functions and BS amplitudes. Based on our current first-principles QCD calculations with fRG \cite{Fu:2025hcm}, we hope to report on the results of PDAs or PDFs of pion or other hadrons in QCD in the near future.

\section*{Acknowledgements}

We thank Lei Chang and Jan M. Pawlowski for discussions. This work is supported by the National Natural Science Foundation of China under Grant Nos.\ 12447102, 12175030, and the Collaborative Research Centre SFB 1225 (ISOQUANT).

\appendix 

\section{Quasi-PDA in the Point-like limit}
\label{app:pointlike}

In the appendix, we would like to derive the pion PDA in the point-like pion and asymptotic limits in the quasi-light-front framework analytically, using the integration shift discussed in \Cref{sec:integ-contour}.

First of all, we begin with the point-like limit. In this limit, the momentum dependence of quark mass and BS amplitude is neglected, and here we take $h_{\pi}=1$. Note that the constant BS amplitude also corresponds to the chiral limit $m_{\pi}=0$, see \cite{Roberts:2010rn}. Then the quasi-LFWF is given by
\begin{align}
    &\psi_\pi (x,P_{z},p_{\perp})\nonumber\\[2ex]
    =&\frac{1}{\pi f_\pi}\mathrm{Tr}_{\mathrm{CD}}\bigg[\int d p_0 d p_3\delta(\tilde n\cdot p_+-x\tilde n\cdot P)\gamma_5 \gamma\cdot \tilde n \chi_\pi(p;P)\bigg]\nonumber\\[2ex]
    =&\frac{4 N_c}{\pi f_\pi}\int d p_0 P_z M_q \frac{1}
{(p_0+\mathrm{i} P_z/2)^2+(x P_z)^2+p_{\perp}^{2}+M_q^2}\nonumber\\[2ex]
    &\times \frac{1}{(p_0-\mathrm{i} P_z/2)^2+(1-x)^2 P_z^2+p_{\perp}^{2}+M_q^2}\,,\label{eq:qLFWF}
\end{align}
where the quark wave function is assumed to be $Z_q=1$. Hence, the poles in \labelcref{eq:poles} are reduced to
\begin{align}
     p_{0,\mathrm{pole}\,1}=&\mathrm{i}\bigg[-\sqrt{p_\perp^2+(xP_z)^2+M_q^2}-\frac{P_z}{2}\bigg]\,,\nonumber\\[2ex]
     p_{0,\mathrm{pole}\,2}=&\mathrm{i}\bigg[\sqrt{p_\perp^2+(xP_z)^2+M_q^2}-\frac{P_z}{2}\bigg]\,,\nonumber\\[2ex]
     p_{0,\mathrm{pole}\,3}=&\mathrm{i}\bigg[-\sqrt{p_\perp^2+\big((1-x)P_z \big)^2+M_q^2}+\frac{P_z}{2}\bigg]\,,\nonumber\\[2ex]
     p_{0,\mathrm{pole}\,4}=&\mathrm{i}\bigg[\sqrt{p_\perp^2+\big((1-x)P_z \big)^2+M_q^2}+\frac{P_z}{2}\bigg]\,.\label{eq:four-poles}
\end{align}
According to the shifted $p_0$-integral in \labelcref{eq:integral-shift} with \labelcref{eq:deform-cont}, it is equivalent to compute the residues at $p_{0,\mathrm{pole}\,1}$ and $p_{0,\mathrm{pole}\,3}$ or $p_{0,\mathrm{pole}\,2}$ and $p_{0,\mathrm{pole}\,4}$. Then one arrives at
\begin{align}
    &\psi_\pi (x,P_{z},p_{\perp})\nonumber\\[2ex]
    =&\frac{2 N_c}{f_\pi} M_q P_z\bigg\{ \Big[ P_z\big(p_\perp^2+(xP_z)^2+M_q^2\big)\nonumber\\[2ex]
    &-xP_z^2\big(p_\perp^2+(xP_z)^2+M_q^2\big)^{1/2}\Big]^{-1}\nonumber\\[2ex]
    &+(1-x)P_z^2\big(p_\perp^2+((1-x)P_z)^2+M_q^2\big)^{1/2}\Big]^{-1} \nonumber\\[2ex]
    &-\Big[ P_z\big(p_\perp^2+((1-x)P_z)^2+M_q^2\big)\bigg\}\,.\label{eq:qLFWF2}
\end{align}
It is not difficult to find that the first term in the curly bracket in \labelcref{eq:qLFWF2} behaves as $\sim P_z^{-1}$ as $P_z \to \infty$ and the second term $\sim P_z^{-3}$. Therefore, the second can be neglected safely, and one finally finds the pion light-front wave function
\begin{align}
    \psi_{\pi}^{\mathrm{LF}} (x,p_{\perp})\equiv\psi_\pi (x,P_{z} \to \infty,p_{\perp})=&\frac{4 N_c}{f_\pi} \frac{M_q}{p_\perp^2+M_q^2}\,.\label{eq:qLFWF3}
\end{align}
Evidently, the LFWF above is independent of the momentum fraction $x$, and so is the PDA, which reads
\begin{align}
    \varphi_{\pi}(x)=1\,,
\end{align}
for the point-like case. In this case, the pion can be regarded as a point-like particle without inner structure.

\section{Quasi-PDA in the asymptotic limit}
\label{app:asymptotic}

The asymptotic limit corresponds to the ultraviolet high energy regime of $p\gg \Lambda_{\mathrm{QCD}}$. In this regime, the quark mass can be treated as a constant, and the pion meson mass can be neglected as such $m_{\pi}=0$. Here we employ the first-order Nakanishi-like representation for the BS amplitude, see also \cite{Chang:2013pq}. Since the PDA would be normalized finally, one can ignore constant prefactors in the BS amplitude as well as the quasi-LFWF in \labelcref{eq:qLFWF} for the moment. Inserting the quasi-light-front BS amplitude as follows
\begin{align}
    &h_{\pi}(p;P)\nonumber\\[2ex]
    =&\int_{-1}^{1}dw\,\frac{1-w^2}{(p_0+\mathrm{i} wP_{z}/2)^2+(p_3+wP_{z}/2)^2+p_{\perp}^2+M_q^2}\,,
\end{align}
one finds for the quasi-LFWF
\begin{align}
    &{\psi}_\pi (x,P_{z},p_{\perp})\nonumber\\[2ex]
    =& M_q P_z\int d p_0\int_{-1}^1 dw (1-w^2)\nonumber\\[2ex]
    &\times\frac{1}{(p_0+\mathrm{i} w P_{z}/2)^2+(w+2x-1)^2 P_z^2/4+p_{\perp}^2+M_q^2}\nonumber\\[2ex]
    &\times \frac{1}{(p_0+\mathrm{i} P_{z}/2)^2+x^2P_z^2+p_{\perp}^2+M_q^2}\nonumber\\[2ex]
    &\times\frac{1}{(p_0-\mathrm{i} P_{z}/2)^2+(1-x)^2P_z^2+p_{\perp}^2+M_q^2}\,.\label{eq:asymptotic-express}
\end{align}
Note that there are two other poles from the BS amplitude besides the four poles in \labelcref{eq:four-poles}, which are denoted by
\begin{align}
     p_{0,\mathrm{pole}\,5}=&\mathrm{i}\bigg[-\sqrt{p_\perp^2+\frac{(w+2x-1)^2P_z^2}{4}+M_q^2}-\frac{w P_{z}}{2}\bigg]\,,\nonumber\\[2ex]
     p_{0,\mathrm{pole}\,6}=&\mathrm{i}\bigg[\sqrt{p_\perp^2+\frac{(w+2x-1)^2P_z^2}{4}+M_q^2}-\frac{w P_{z}}{2}\bigg]\,.\label{}
\end{align}
One is able to integrate $p_0$ in \labelcref{eq:asymptotic-express} as before by computing the residues at $(p_{0,\mathrm{pole}\,1},\,p_{0,\mathrm{pole}\,3},p_{0,\mathrm{pole}\,5}\,)$ or $(p_{0,\mathrm{pole}\,2},\,p_{0,\mathrm{pole}\,4},p_{0,\mathrm{pole}\,6}\,)$. Then, the result is given by
\begin{widetext}
\begin{align}
    \psi_\pi (x,P_{z},p_{\perp})=&\pi M_q P_z\int_{-1}^1 dw \Bigg[\frac{1+w}{2}\frac{1}{\sqrt{p_\perp^2+(xP_z)^2+M_q^2}}\frac{1}{\Big(P_z\sqrt{p_\perp^2+(xP_z)^2+M_q^2}-xP_z^2\Big)^2}\nonumber\\[2ex]
    &+\frac{1-w}{2}\frac{1}{\sqrt{p_\perp^2+\big((1-x)P_z \big)^2+M_q^2}}\frac{1}{\Big(P_z\sqrt{p_\perp^2+\big((1-x)P_z \big)^2+M_q^2}+(1-x)P_z^2\Big)^2}\nonumber\\[2ex]
    &-\frac{1}{\sqrt{p_\perp^2+(w+2x-1)^2P_z^2/4+M_q^2}}\frac{1}{\Big(P_z\sqrt{p_\perp^2+(w+2x-1)^2P_z^2/4+M_q^2}-(w+2x-1)P_z^2/2\Big)^2}\Bigg]\,.\label{eq:qLFWF-asymp}
\end{align}
\end{widetext}
It is not difficult to find that the first term in the square bracket in \labelcref{eq:qLFWF-asymp} has the leading behavior $\sim P_z^{-1}$ in the limit of $P_z \to \infty$, the second term $\sim P_z^{-5}$ that would not contribute to the light-front PDA. For the third term, we have to distinguish two cases: One is $w+2x-1>0$ and it has the leading behavior $\sim P_z^{-1}$ that contributes to PDA. The other is $w+2x-1<0$, which leads to $\sim P_z^{-5}$ and thus can be neglected. Taking all these analyses into account, one is led to
\begin{align}
    \psi_{\pi}^{\mathrm{LF}}&(x,p_{\perp})  =4 \pi\frac{M_q}{(p_\perp^2+M_q^2)^2}\nonumber\\[2ex]&\times\Big(\int_{-1}^1 dw \frac{1+w}{2}x
    -\int_{1-2x}^1 dw \frac{2x-1+w}{2}\Big)\,.\label{}
\end{align}
Finally, we arrive
\begin{align}
    \psi_{\pi}^{\mathrm{LF}} (x,p_{\perp})=4 \pi\frac{M_q}{(p_\perp^2+M_q^2)^2}x(1-x)\,.
\end{align}
In consequence, the light-front PDA of the pion in the asymptotic limit reads
\begin{align}
    \varphi_{\pi}(x)=6x(1-x)\,,
\end{align}
which has been normalized.

\vfill 

\bibliography{ref-lib}%

\begin{thebibliography}{55}%
\makeatletter
\providecommand \@ifxundefined [1]{%
 \@ifx{#1\undefined}
}%
\providecommand \@ifnum [1]{%
 \ifnum #1\expandafter \@firstoftwo
 \else \expandafter \@secondoftwo
 \fi
}%
\providecommand \@ifx [1]{%
 \ifx #1\expandafter \@firstoftwo
 \else \expandafter \@secondoftwo
 \fi
}%
\providecommand \natexlab [1]{#1}%
\providecommand \enquote  [1]{``#1''}%
\providecommand \bibnamefont  [1]{#1}%
\providecommand \bibfnamefont [1]{#1}%
\providecommand \citenamefont [1]{#1}%
\providecommand \href@noop [0]{\@secondoftwo}%
\providecommand \href [0]{\begingroup \@sanitize@url \@href}%
\providecommand \@href[1]{\@@startlink{#1}\@@href}%
\providecommand \@@href[1]{\endgroup#1\@@endlink}%
\providecommand \@sanitize@url [0]{\catcode `\\12\catcode `\$12\catcode
  `\&12\catcode `\#12\catcode `\^12\catcode `\_12\catcode `\%12\relax}%
\providecommand \@@startlink[1]{}%
\providecommand \@@endlink[0]{}%
\providecommand \url  [0]{\begingroup\@sanitize@url \@url }%
\providecommand \@url [1]{\endgroup\@href {#1}{\urlprefix }}%
\providecommand \urlprefix  [0]{URL }%
\providecommand \Eprint [0]{\href }%
\providecommand \doibase [0]{https://doi.org/}%
\providecommand \selectlanguage [0]{\@gobble}%
\providecommand \bibinfo  [0]{\@secondoftwo}%
\providecommand \bibfield  [0]{\@secondoftwo}%
\providecommand \translation [1]{[#1]}%
\providecommand \BibitemOpen [0]{}%
\providecommand \bibitemStop [0]{}%
\providecommand \bibitemNoStop [0]{.\EOS\space}%
\providecommand \EOS [0]{\spacefactor3000\relax}%
\providecommand \BibitemShut  [1]{\csname bibitem#1\endcsname}%
\let\auto@bib@innerbib\@empty
\bibitem [{\citenamefont {Marciano}\ and\ \citenamefont
  {Pagels}(1978)}]{Marciano:1977su}%
  \BibitemOpen
  \bibfield  {author} {\bibinfo {author} {\bibfnamefont {W.~J.}\ \bibnamefont
  {Marciano}}\ and\ \bibinfo {author} {\bibfnamefont {H.}~\bibnamefont
  {Pagels}},\ }\bibfield  {title} {\bibinfo {title} {{Quantum Chromodynamics: A
  Review}},\ }\href {https://doi.org/10.1016/0370-1573(78)90208-9} {\bibfield
  {journal} {\bibinfo  {journal} {Phys. Rept.}\ }\textbf {\bibinfo {volume}
  {36}},\ \bibinfo {pages} {137} (\bibinfo {year} {1978})}\BibitemShut
  {NoStop}%
\bibitem [{\citenamefont {Marciano}\ and\ \citenamefont
  {Pagels}(1979)}]{Marciano:1979wa}%
  \BibitemOpen
  \bibfield  {author} {\bibinfo {author} {\bibfnamefont {W.~J.}\ \bibnamefont
  {Marciano}}\ and\ \bibinfo {author} {\bibfnamefont {H.}~\bibnamefont
  {Pagels}},\ }\bibfield  {title} {\bibinfo {title} {{QUANTUM
  CHROMODYNAMICS}},\ }\href {https://doi.org/10.1038/279479a0} {\bibfield
  {journal} {\bibinfo  {journal} {Nature}\ }\textbf {\bibinfo {volume} {279}},\
  \bibinfo {pages} {479} (\bibinfo {year} {1979})}\BibitemShut {NoStop}%
\bibitem [{\citenamefont {Cheng}\ and\ \citenamefont
  {Chua}(2009)}]{Cheng:2009cn}%
  \BibitemOpen
  \bibfield  {author} {\bibinfo {author} {\bibfnamefont {H.-Y.}\ \bibnamefont
  {Cheng}}\ and\ \bibinfo {author} {\bibfnamefont {C.-K.}\ \bibnamefont
  {Chua}},\ }\bibfield  {title} {\bibinfo {title} {{Revisiting Charmless
  Hadronic B(u,d) Decays in QCD Factorization}},\ }\href
  {https://doi.org/10.1103/PhysRevD.80.114008} {\bibfield  {journal} {\bibinfo
  {journal} {Phys. Rev. D}\ }\textbf {\bibinfo {volume} {80}},\ \bibinfo
  {pages} {114008} (\bibinfo {year} {2009})},\ \Eprint
  {https://arxiv.org/abs/0909.5229} {arXiv:0909.5229 [hep-ph]} \BibitemShut
  {NoStop}%
\bibitem [{\citenamefont {Su}\ \emph {et~al.}(2011)\citenamefont {Su},
  \citenamefont {Wu}, \citenamefont {Yang},\ and\ \citenamefont
  {Zhuang}}]{Su:2010vt}%
  \BibitemOpen
  \bibfield  {author} {\bibinfo {author} {\bibfnamefont {F.}~\bibnamefont
  {Su}}, \bibinfo {author} {\bibfnamefont {Y.-L.}\ \bibnamefont {Wu}}, \bibinfo
  {author} {\bibfnamefont {Y.-B.}\ \bibnamefont {Yang}},\ and\ \bibinfo
  {author} {\bibfnamefont {C.}~\bibnamefont {Zhuang}},\ }\bibfield  {title}
  {\bibinfo {title} {{Charmless $B\to PP, PV, VV$ Decays Based on the six-quark
  Effective Hamiltonian with Strong Phase Effects I}},\ }\href
  {https://doi.org/10.1088/0954-3899/38/1/015006} {\bibfield  {journal}
  {\bibinfo  {journal} {J. Phys. G}\ }\textbf {\bibinfo {volume} {38}},\
  \bibinfo {pages} {015006} (\bibinfo {year} {2011})},\ \Eprint
  {https://arxiv.org/abs/1006.1100} {arXiv:1006.1100 [hep-ph]} \BibitemShut
  {NoStop}%
\bibitem [{\citenamefont {Aaij}\ \emph {et~al.}(2019)\citenamefont {Aaij} \emph
  {et~al.}}]{LHCb:2019hip}%
  \BibitemOpen
  \bibfield  {author} {\bibinfo {author} {\bibfnamefont {R.}~\bibnamefont
  {Aaij}} \emph {et~al.} (\bibinfo {collaboration} {LHCb}),\ }\bibfield
  {title} {\bibinfo {title} {{Search for lepton-universality violation in
  $B^+\to K^+\ell^+\ell^-$ decays}},\ }\href
  {https://doi.org/10.1103/PhysRevLett.122.191801} {\bibfield  {journal}
  {\bibinfo  {journal} {Phys. Rev. Lett.}\ }\textbf {\bibinfo {volume} {122}},\
  \bibinfo {pages} {191801} (\bibinfo {year} {2019})},\ \Eprint
  {https://arxiv.org/abs/1903.09252} {arXiv:1903.09252 [hep-ex]} \BibitemShut
  {NoStop}%
\bibitem [{\citenamefont {Berger}\ and\ \citenamefont
  {Brodsky}(1979)}]{Berger:1979du}%
  \BibitemOpen
  \bibfield  {author} {\bibinfo {author} {\bibfnamefont {E.~L.}\ \bibnamefont
  {Berger}}\ and\ \bibinfo {author} {\bibfnamefont {S.~J.}\ \bibnamefont
  {Brodsky}},\ }\bibfield  {title} {\bibinfo {title} {{Quark Structure
  Functions of Mesons and the Drell-Yan Process}},\ }\href
  {https://doi.org/10.1103/PhysRevLett.42.940} {\bibfield  {journal} {\bibinfo
  {journal} {Phys. Rev. Lett.}\ }\textbf {\bibinfo {volume} {42}},\ \bibinfo
  {pages} {940} (\bibinfo {year} {1979})}\BibitemShut {NoStop}%
\bibitem [{\citenamefont {Badier}\ \emph {et~al.}(1980)\citenamefont {Badier}
  \emph {et~al.}}]{Saclay-CERN-CollegedeFrance-EcolePoly-Orsay:1980fhh}%
  \BibitemOpen
  \bibfield  {author} {\bibinfo {author} {\bibfnamefont {J.}~\bibnamefont
  {Badier}} \emph {et~al.} (\bibinfo {collaboration} {Saclay-CERN-College de
  France-Ecole Poly-Orsay}),\ }\bibfield  {title} {\bibinfo {title}
  {{Measurement of the $K^- / \pi^-$ Structure Function Ratio Using the
  {Drell-Yan} Process}},\ }\href {https://doi.org/10.1016/0370-2693(80)90530-4}
  {\bibfield  {journal} {\bibinfo  {journal} {Phys. Lett. B}\ }\textbf
  {\bibinfo {volume} {93}},\ \bibinfo {pages} {354} (\bibinfo {year}
  {1980})}\BibitemShut {NoStop}%
\bibitem [{\citenamefont {Badier}\ \emph {et~al.}(1983)\citenamefont {Badier}
  \emph {et~al.}}]{NA3:1983ejh}%
  \BibitemOpen
  \bibfield  {author} {\bibinfo {author} {\bibfnamefont {J.}~\bibnamefont
  {Badier}} \emph {et~al.} (\bibinfo {collaboration} {NA3}),\ }\bibfield
  {title} {\bibinfo {title} {{Experimental Determination of the pi Meson
  Structure Functions by the Drell-Yan Mechanism}},\ }\href
  {https://doi.org/10.1007/BF01573728} {\bibfield  {journal} {\bibinfo
  {journal} {Z. Phys. C}\ }\textbf {\bibinfo {volume} {18}},\ \bibinfo {pages}
  {281} (\bibinfo {year} {1983})}\BibitemShut {NoStop}%
\bibitem [{\citenamefont {Aguilar}\ \emph {et~al.}(2019)\citenamefont {Aguilar}
  \emph {et~al.}}]{Aguilar:2019teb}%
  \BibitemOpen
  \bibfield  {author} {\bibinfo {author} {\bibfnamefont {A.~C.}\ \bibnamefont
  {Aguilar}} \emph {et~al.},\ }\bibfield  {title} {\bibinfo {title} {{Pion and
  Kaon Structure at the Electron-Ion Collider}},\ }\href
  {https://doi.org/10.1140/epja/i2019-12885-0} {\bibfield  {journal} {\bibinfo
  {journal} {Eur. Phys. J. A}\ }\textbf {\bibinfo {volume} {55}},\ \bibinfo
  {pages} {190} (\bibinfo {year} {2019})},\ \Eprint
  {https://arxiv.org/abs/1907.08218} {arXiv:1907.08218 [nucl-ex]} \BibitemShut
  {NoStop}%
\bibitem [{\citenamefont {Anderle}\ \emph {et~al.}(2021)\citenamefont {Anderle}
  \emph {et~al.}}]{Anderle:2021wcy}%
  \BibitemOpen
  \bibfield  {author} {\bibinfo {author} {\bibfnamefont {D.~P.}\ \bibnamefont
  {Anderle}} \emph {et~al.},\ }\bibfield  {title} {\bibinfo {title}
  {{Electron-ion collider in China}},\ }\href
  {https://doi.org/10.1007/s11467-021-1062-0} {\bibfield  {journal} {\bibinfo
  {journal} {Front. Phys. (Beijing)}\ }\textbf {\bibinfo {volume} {16}},\
  \bibinfo {pages} {64701} (\bibinfo {year} {2021})},\ \Eprint
  {https://arxiv.org/abs/2102.09222} {arXiv:2102.09222 [nucl-ex]} \BibitemShut
  {NoStop}%
\bibitem [{\citenamefont {Abir}\ \emph {et~al.}(2023)\citenamefont {Abir} \emph
  {et~al.}}]{Abir:2023fpo}%
  \BibitemOpen
  \bibfield  {author} {\bibinfo {author} {\bibfnamefont {R.}~\bibnamefont
  {Abir}} \emph {et~al.},\ }\bibfield  {title} {\bibinfo {title} {{The case for
  an EIC Theory Alliance: Theoretical Challenges of the EIC}},\ }\href@noop {}
  {\  (\bibinfo {year} {2023})},\ \Eprint {https://arxiv.org/abs/2305.14572}
  {arXiv:2305.14572 [hep-ph]} \BibitemShut {NoStop}%
\bibitem [{\citenamefont {Achenbach}\ \emph {et~al.}(2024)\citenamefont
  {Achenbach} \emph {et~al.}}]{Achenbach:2023pba}%
  \BibitemOpen
  \bibfield  {author} {\bibinfo {author} {\bibfnamefont {P.}~\bibnamefont
  {Achenbach}} \emph {et~al.},\ }\bibfield  {title} {\bibinfo {title} {{The
  present and future of QCD}},\ }\href
  {https://doi.org/10.1016/j.nuclphysa.2024.122874} {\bibfield  {journal}
  {\bibinfo  {journal} {Nucl. Phys. A}\ }\textbf {\bibinfo {volume} {1047}},\
  \bibinfo {pages} {122874} (\bibinfo {year} {2024})},\ \Eprint
  {https://arxiv.org/abs/2303.02579} {arXiv:2303.02579 [hep-ph]} \BibitemShut
  {NoStop}%
\bibitem [{\citenamefont {Eichmann}\ \emph {et~al.}(2022)\citenamefont
  {Eichmann}, \citenamefont {Ferreira},\ and\ \citenamefont
  {Stadler}}]{Eichmann:2021vnj}%
  \BibitemOpen
  \bibfield  {author} {\bibinfo {author} {\bibfnamefont {G.}~\bibnamefont
  {Eichmann}}, \bibinfo {author} {\bibfnamefont {E.}~\bibnamefont {Ferreira}},\
  and\ \bibinfo {author} {\bibfnamefont {A.}~\bibnamefont {Stadler}},\
  }\bibfield  {title} {\bibinfo {title} {{Going to the light front with contour
  deformations}},\ }\href {https://doi.org/10.1103/PhysRevD.105.034009}
  {\bibfield  {journal} {\bibinfo  {journal} {Phys. Rev. D}\ }\textbf {\bibinfo
  {volume} {105}},\ \bibinfo {pages} {034009} (\bibinfo {year} {2022})},\
  \Eprint {https://arxiv.org/abs/2112.04858} {arXiv:2112.04858 [hep-ph]}
  \BibitemShut {NoStop}%
\bibitem [{\citenamefont {Arthur}\ \emph {et~al.}(2011)\citenamefont {Arthur},
  \citenamefont {Boyle}, \citenamefont {Brommel}, \citenamefont {Donnellan},
  \citenamefont {Flynn}, \citenamefont {Juttner}, \citenamefont {Rae},\ and\
  \citenamefont {Sachrajda}}]{Arthur:2010xf}%
  \BibitemOpen
  \bibfield  {author} {\bibinfo {author} {\bibfnamefont {R.}~\bibnamefont
  {Arthur}}, \bibinfo {author} {\bibfnamefont {P.~A.}\ \bibnamefont {Boyle}},
  \bibinfo {author} {\bibfnamefont {D.}~\bibnamefont {Brommel}}, \bibinfo
  {author} {\bibfnamefont {M.~A.}\ \bibnamefont {Donnellan}}, \bibinfo {author}
  {\bibfnamefont {J.~M.}\ \bibnamefont {Flynn}}, \bibinfo {author}
  {\bibfnamefont {A.}~\bibnamefont {Juttner}}, \bibinfo {author} {\bibfnamefont
  {T.~D.}\ \bibnamefont {Rae}},\ and\ \bibinfo {author} {\bibfnamefont
  {C.~T.~C.}\ \bibnamefont {Sachrajda}},\ }\bibfield  {title} {\bibinfo {title}
  {{Lattice Results for Low Moments of Light Meson Distribution Amplitudes}},\
  }\href {https://doi.org/10.1103/PhysRevD.83.074505} {\bibfield  {journal}
  {\bibinfo  {journal} {Phys. Rev. D}\ }\textbf {\bibinfo {volume} {83}},\
  \bibinfo {pages} {074505} (\bibinfo {year} {2011})},\ \Eprint
  {https://arxiv.org/abs/1011.5906} {arXiv:1011.5906 [hep-lat]} \BibitemShut
  {NoStop}%
\bibitem [{\citenamefont {Braun}\ \emph {et~al.}(2015)\citenamefont {Braun},
  \citenamefont {Collins}, \citenamefont {G\"ockeler}, \citenamefont
  {P\'erez-Rubio}, \citenamefont {Sch\"afer}, \citenamefont {Schiel},\ and\
  \citenamefont {Sternbeck}}]{Braun:2015axa}%
  \BibitemOpen
  \bibfield  {author} {\bibinfo {author} {\bibfnamefont {V.~M.}\ \bibnamefont
  {Braun}}, \bibinfo {author} {\bibfnamefont {S.}~\bibnamefont {Collins}},
  \bibinfo {author} {\bibfnamefont {M.}~\bibnamefont {G\"ockeler}}, \bibinfo
  {author} {\bibfnamefont {P.}~\bibnamefont {P\'erez-Rubio}}, \bibinfo {author}
  {\bibfnamefont {A.}~\bibnamefont {Sch\"afer}}, \bibinfo {author}
  {\bibfnamefont {R.~W.}\ \bibnamefont {Schiel}},\ and\ \bibinfo {author}
  {\bibfnamefont {A.}~\bibnamefont {Sternbeck}},\ }\bibfield  {title} {\bibinfo
  {title} {{Second Moment of the Pion Light-cone Distribution Amplitude from
  Lattice QCD}},\ }\href {https://doi.org/10.1103/PhysRevD.92.014504}
  {\bibfield  {journal} {\bibinfo  {journal} {Phys. Rev. D}\ }\textbf {\bibinfo
  {volume} {92}},\ \bibinfo {pages} {014504} (\bibinfo {year} {2015})},\
  \Eprint {https://arxiv.org/abs/1503.03656} {arXiv:1503.03656 [hep-lat]}
  \BibitemShut {NoStop}%
\bibitem [{\citenamefont {Bali}\ \emph {et~al.}(2017)\citenamefont {Bali},
  \citenamefont {Braun}, \citenamefont {G\"ockeler}, \citenamefont {Gruber},
  \citenamefont {Hutzler}, \citenamefont {Korcyl}, \citenamefont {Lang},\ and\
  \citenamefont {Sch\"afer}}]{Bali:2017ude}%
  \BibitemOpen
  \bibfield  {author} {\bibinfo {author} {\bibfnamefont {G.~S.}\ \bibnamefont
  {Bali}}, \bibinfo {author} {\bibfnamefont {V.~M.}\ \bibnamefont {Braun}},
  \bibinfo {author} {\bibfnamefont {M.}~\bibnamefont {G\"ockeler}}, \bibinfo
  {author} {\bibfnamefont {M.}~\bibnamefont {Gruber}}, \bibinfo {author}
  {\bibfnamefont {F.}~\bibnamefont {Hutzler}}, \bibinfo {author} {\bibfnamefont
  {P.}~\bibnamefont {Korcyl}}, \bibinfo {author} {\bibfnamefont
  {B.}~\bibnamefont {Lang}},\ and\ \bibinfo {author} {\bibfnamefont
  {A.}~\bibnamefont {Sch\"afer}} (\bibinfo {collaboration} {RQCD}),\ }\bibfield
   {title} {\bibinfo {title} {{Second moment of the pion distribution amplitude
  with the momentum smearing technique}},\ }\href
  {https://doi.org/10.1016/j.physletb.2017.08.077} {\bibfield  {journal}
  {\bibinfo  {journal} {Phys. Lett. B}\ }\textbf {\bibinfo {volume} {774}},\
  \bibinfo {pages} {91} (\bibinfo {year} {2017})},\ \Eprint
  {https://arxiv.org/abs/1705.10236} {arXiv:1705.10236 [hep-lat]} \BibitemShut
  {NoStop}%
\bibitem [{\citenamefont {Bali}\ \emph {et~al.}(2019)\citenamefont {Bali},
  \citenamefont {Braun}, \citenamefont {B\"urger}, \citenamefont {G\"ockeler},
  \citenamefont {Gruber}, \citenamefont {Hutzler}, \citenamefont {Korcyl},
  \citenamefont {Sch\"afer}, \citenamefont {Sternbeck},\ and\ \citenamefont
  {Wein}}]{RQCD:2019osh}%
  \BibitemOpen
  \bibfield  {author} {\bibinfo {author} {\bibfnamefont {G.~S.}\ \bibnamefont
  {Bali}}, \bibinfo {author} {\bibfnamefont {V.~M.}\ \bibnamefont {Braun}},
  \bibinfo {author} {\bibfnamefont {S.}~\bibnamefont {B\"urger}}, \bibinfo
  {author} {\bibfnamefont {M.}~\bibnamefont {G\"ockeler}}, \bibinfo {author}
  {\bibfnamefont {M.}~\bibnamefont {Gruber}}, \bibinfo {author} {\bibfnamefont
  {F.}~\bibnamefont {Hutzler}}, \bibinfo {author} {\bibfnamefont
  {P.}~\bibnamefont {Korcyl}}, \bibinfo {author} {\bibfnamefont
  {A.}~\bibnamefont {Sch\"afer}}, \bibinfo {author} {\bibfnamefont
  {A.}~\bibnamefont {Sternbeck}},\ and\ \bibinfo {author} {\bibfnamefont
  {P.}~\bibnamefont {Wein}} (\bibinfo {collaboration} {RQCD}),\ }\bibfield
  {title} {\bibinfo {title} {{Light-cone distribution amplitudes of
  pseudoscalar mesons from lattice QCD}},\ }\href
  {https://doi.org/10.1007/JHEP08(2019)065} {\bibfield  {journal} {\bibinfo
  {journal} {JHEP}\ }\textbf {\bibinfo {volume} {08}},\ \bibinfo {pages}
  {065}},\ \bibinfo {note} {[Addendum: JHEP 11, 037 (2020)]},\ \Eprint
  {https://arxiv.org/abs/1903.08038} {arXiv:1903.08038 [hep-lat]} \BibitemShut
  {NoStop}%
\bibitem [{\citenamefont {L\"offler}\ \emph {et~al.}(2022)\citenamefont
  {L\"offler}, \citenamefont {Wein}, \citenamefont {Wurm}, \citenamefont
  {Weish\"aupl}, \citenamefont {Jenkins}, \citenamefont {R\"odl}, \citenamefont
  {Sch\"afer},\ and\ \citenamefont {Walter}}]{Loffler:2021afv}%
  \BibitemOpen
  \bibfield  {author} {\bibinfo {author} {\bibfnamefont {M.}~\bibnamefont
  {L\"offler}}, \bibinfo {author} {\bibfnamefont {P.}~\bibnamefont {Wein}},
  \bibinfo {author} {\bibfnamefont {T.}~\bibnamefont {Wurm}}, \bibinfo {author}
  {\bibfnamefont {S.}~\bibnamefont {Weish\"aupl}}, \bibinfo {author}
  {\bibfnamefont {D.}~\bibnamefont {Jenkins}}, \bibinfo {author} {\bibfnamefont
  {R.}~\bibnamefont {R\"odl}}, \bibinfo {author} {\bibfnamefont
  {A.}~\bibnamefont {Sch\"afer}},\ and\ \bibinfo {author} {\bibfnamefont
  {L.}~\bibnamefont {Walter}} (\bibinfo {collaboration} {RQCD}),\ }\bibfield
  {title} {\bibinfo {title} {{Mellin moments of spin dependent and independent
  PDFs of the pion and rho meson}},\ }\href
  {https://doi.org/10.1103/PhysRevD.105.014505} {\bibfield  {journal} {\bibinfo
   {journal} {Phys. Rev. D}\ }\textbf {\bibinfo {volume} {105}},\ \bibinfo
  {pages} {014505} (\bibinfo {year} {2022})},\ \Eprint
  {https://arxiv.org/abs/2108.07544} {arXiv:2108.07544 [hep-lat]} \BibitemShut
  {NoStop}%
\bibitem [{\citenamefont {Chang}\ \emph {et~al.}(2013)\citenamefont {Chang},
  \citenamefont {Cloet}, \citenamefont {Cobos-Martinez}, \citenamefont
  {Roberts}, \citenamefont {Schmidt},\ and\ \citenamefont
  {Tandy}}]{Chang:2013pq}%
  \BibitemOpen
  \bibfield  {author} {\bibinfo {author} {\bibfnamefont {L.}~\bibnamefont
  {Chang}}, \bibinfo {author} {\bibfnamefont {I.~C.}\ \bibnamefont {Cloet}},
  \bibinfo {author} {\bibfnamefont {J.~J.}\ \bibnamefont {Cobos-Martinez}},
  \bibinfo {author} {\bibfnamefont {C.~D.}\ \bibnamefont {Roberts}}, \bibinfo
  {author} {\bibfnamefont {S.~M.}\ \bibnamefont {Schmidt}},\ and\ \bibinfo
  {author} {\bibfnamefont {P.~C.}\ \bibnamefont {Tandy}},\ }\bibfield  {title}
  {\bibinfo {title} {{Imaging dynamical chiral symmetry breaking: pion wave
  function on the light front}},\ }\href
  {https://doi.org/10.1103/PhysRevLett.110.132001} {\bibfield  {journal}
  {\bibinfo  {journal} {Phys. Rev. Lett.}\ }\textbf {\bibinfo {volume} {110}},\
  \bibinfo {pages} {132001} (\bibinfo {year} {2013})},\ \Eprint
  {https://arxiv.org/abs/1301.0324} {arXiv:1301.0324 [nucl-th]} \BibitemShut
  {NoStop}%
\bibitem [{\citenamefont {Chang}\ \emph {et~al.}(2014)\citenamefont {Chang},
  \citenamefont {Mezrag}, \citenamefont {Moutarde}, \citenamefont {Roberts},
  \citenamefont {Rodr\'\i{}guez-Quintero},\ and\ \citenamefont
  {Tandy}}]{Chang:2014lva}%
  \BibitemOpen
  \bibfield  {author} {\bibinfo {author} {\bibfnamefont {L.}~\bibnamefont
  {Chang}}, \bibinfo {author} {\bibfnamefont {C.}~\bibnamefont {Mezrag}},
  \bibinfo {author} {\bibfnamefont {H.}~\bibnamefont {Moutarde}}, \bibinfo
  {author} {\bibfnamefont {C.~D.}\ \bibnamefont {Roberts}}, \bibinfo {author}
  {\bibfnamefont {J.}~\bibnamefont {Rodr\'\i{}guez-Quintero}},\ and\ \bibinfo
  {author} {\bibfnamefont {P.~C.}\ \bibnamefont {Tandy}},\ }\bibfield  {title}
  {\bibinfo {title} {{Basic features of the pion valence-quark distribution
  function}},\ }\href {https://doi.org/10.1016/j.physletb.2014.08.009}
  {\bibfield  {journal} {\bibinfo  {journal} {Phys. Lett. B}\ }\textbf
  {\bibinfo {volume} {737}},\ \bibinfo {pages} {23} (\bibinfo {year} {2014})},\
  \Eprint {https://arxiv.org/abs/1406.5450} {arXiv:1406.5450 [nucl-th]}
  \BibitemShut {NoStop}%
\bibitem [{\citenamefont {Chen}\ \emph {et~al.}(2016)\citenamefont {Chen},
  \citenamefont {Chang}, \citenamefont {Roberts}, \citenamefont {Wan},\ and\
  \citenamefont {Zong}}]{Chen:2016sno}%
  \BibitemOpen
  \bibfield  {author} {\bibinfo {author} {\bibfnamefont {C.}~\bibnamefont
  {Chen}}, \bibinfo {author} {\bibfnamefont {L.}~\bibnamefont {Chang}},
  \bibinfo {author} {\bibfnamefont {C.~D.}\ \bibnamefont {Roberts}}, \bibinfo
  {author} {\bibfnamefont {S.}~\bibnamefont {Wan}},\ and\ \bibinfo {author}
  {\bibfnamefont {H.-S.}\ \bibnamefont {Zong}},\ }\bibfield  {title} {\bibinfo
  {title} {{Valence-quark distribution functions in the kaon and pion}},\
  }\href {https://doi.org/10.1103/PhysRevD.93.074021} {\bibfield  {journal}
  {\bibinfo  {journal} {Phys. Rev. D}\ }\textbf {\bibinfo {volume} {93}},\
  \bibinfo {pages} {074021} (\bibinfo {year} {2016})},\ \Eprint
  {https://arxiv.org/abs/1602.01502} {arXiv:1602.01502 [nucl-th]} \BibitemShut
  {NoStop}%
\bibitem [{\citenamefont {Ding}\ \emph {et~al.}(2020)\citenamefont {Ding},
  \citenamefont {Raya}, \citenamefont {Binosi}, \citenamefont {Chang},
  \citenamefont {Roberts},\ and\ \citenamefont {Schmidt}}]{Ding:2019lwe}%
  \BibitemOpen
  \bibfield  {author} {\bibinfo {author} {\bibfnamefont {M.}~\bibnamefont
  {Ding}}, \bibinfo {author} {\bibfnamefont {K.}~\bibnamefont {Raya}}, \bibinfo
  {author} {\bibfnamefont {D.}~\bibnamefont {Binosi}}, \bibinfo {author}
  {\bibfnamefont {L.}~\bibnamefont {Chang}}, \bibinfo {author} {\bibfnamefont
  {C.~D.}\ \bibnamefont {Roberts}},\ and\ \bibinfo {author} {\bibfnamefont
  {S.~M.}\ \bibnamefont {Schmidt}},\ }\bibfield  {title} {\bibinfo {title}
  {{Symmetry, symmetry breaking, and pion parton distributions}},\ }\href
  {https://doi.org/10.1103/PhysRevD.101.054014} {\bibfield  {journal} {\bibinfo
   {journal} {Phys. Rev. D}\ }\textbf {\bibinfo {volume} {101}},\ \bibinfo
  {pages} {054014} (\bibinfo {year} {2020})},\ \Eprint
  {https://arxiv.org/abs/1905.05208} {arXiv:1905.05208 [nucl-th]} \BibitemShut
  {NoStop}%
\bibitem [{\citenamefont {Cui}\ \emph {et~al.}(2020)\citenamefont {Cui},
  \citenamefont {Ding}, \citenamefont {Gao}, \citenamefont {Raya},
  \citenamefont {Binosi}, \citenamefont {Chang}, \citenamefont {Roberts},
  \citenamefont {Rodr\'\i{}guez-Quintero},\ and\ \citenamefont
  {Schmidt}}]{Cui:2020tdf}%
  \BibitemOpen
  \bibfield  {author} {\bibinfo {author} {\bibfnamefont {Z.-F.}\ \bibnamefont
  {Cui}}, \bibinfo {author} {\bibfnamefont {M.}~\bibnamefont {Ding}}, \bibinfo
  {author} {\bibfnamefont {F.}~\bibnamefont {Gao}}, \bibinfo {author}
  {\bibfnamefont {K.}~\bibnamefont {Raya}}, \bibinfo {author} {\bibfnamefont
  {D.}~\bibnamefont {Binosi}}, \bibinfo {author} {\bibfnamefont
  {L.}~\bibnamefont {Chang}}, \bibinfo {author} {\bibfnamefont {C.~D.}\
  \bibnamefont {Roberts}}, \bibinfo {author} {\bibfnamefont {J.}~\bibnamefont
  {Rodr\'\i{}guez-Quintero}},\ and\ \bibinfo {author} {\bibfnamefont {S.~M.}\
  \bibnamefont {Schmidt}},\ }\bibfield  {title} {\bibinfo {title} {{Kaon and
  pion parton distributions}},\ }\href
  {https://doi.org/10.1140/epjc/s10052-020-08578-4} {\bibfield  {journal}
  {\bibinfo  {journal} {Eur. Phys. J. C}\ }\textbf {\bibinfo {volume} {80}},\
  \bibinfo {pages} {1064} (\bibinfo {year} {2020})}\BibitemShut {NoStop}%
\bibitem [{\citenamefont {Wang}\ \emph {et~al.}(2025)\citenamefont {Wang},
  \citenamefont {Xing}, \citenamefont {Chang}, \citenamefont {Ding},
  \citenamefont {Raya},\ and\ \citenamefont {Roberts}}]{Wang:2024fjt}%
  \BibitemOpen
  \bibfield  {author} {\bibinfo {author} {\bibfnamefont {X.}~\bibnamefont
  {Wang}}, \bibinfo {author} {\bibfnamefont {Z.}~\bibnamefont {Xing}}, \bibinfo
  {author} {\bibfnamefont {L.}~\bibnamefont {Chang}}, \bibinfo {author}
  {\bibfnamefont {M.}~\bibnamefont {Ding}}, \bibinfo {author} {\bibfnamefont
  {K.}~\bibnamefont {Raya}},\ and\ \bibinfo {author} {\bibfnamefont {C.~D.}\
  \bibnamefont {Roberts}},\ }\bibfield  {title} {\bibinfo {title} {{Sketching
  pion and proton mass distributions}},\ }\href
  {https://doi.org/10.1016/j.physletb.2025.139280} {\bibfield  {journal}
  {\bibinfo  {journal} {Phys. Lett. B}\ }\textbf {\bibinfo {volume} {862}},\
  \bibinfo {pages} {139280} (\bibinfo {year} {2025})},\ \Eprint
  {https://arxiv.org/abs/2410.13068} {arXiv:2410.13068 [hep-ph]} \BibitemShut
  {NoStop}%
\bibitem [{\citenamefont {Horn}\ and\ \citenamefont
  {Roberts}(2016)}]{Horn:2016rip}%
  \BibitemOpen
  \bibfield  {author} {\bibinfo {author} {\bibfnamefont {T.}~\bibnamefont
  {Horn}}\ and\ \bibinfo {author} {\bibfnamefont {C.~D.}\ \bibnamefont
  {Roberts}},\ }\bibfield  {title} {\bibinfo {title} {{The pion: an enigma
  within the Standard Model}},\ }\href
  {https://doi.org/10.1088/0954-3899/43/7/073001} {\bibfield  {journal}
  {\bibinfo  {journal} {J. Phys. G}\ }\textbf {\bibinfo {volume} {43}},\
  \bibinfo {pages} {073001} (\bibinfo {year} {2016})},\ \Eprint
  {https://arxiv.org/abs/1602.04016} {arXiv:1602.04016 [nucl-th]} \BibitemShut
  {NoStop}%
\bibitem [{\citenamefont {Yu}\ and\ \citenamefont
  {Roberts}(2024)}]{Yu:2024ovn}%
  \BibitemOpen
  \bibfield  {author} {\bibinfo {author} {\bibfnamefont {Y.}~\bibnamefont
  {Yu}}\ and\ \bibinfo {author} {\bibfnamefont {C.~D.}\ \bibnamefont
  {Roberts}},\ }\bibfield  {title} {\bibinfo {title} {{Impressions of Parton
  Distribution Functions}},\ }\href
  {https://doi.org/10.1088/0256-307X/41/12/121202} {\bibfield  {journal}
  {\bibinfo  {journal} {Chin. Phys. Lett.}\ }\textbf {\bibinfo {volume} {41}},\
  \bibinfo {pages} {121202} (\bibinfo {year} {2024})},\ \Eprint
  {https://arxiv.org/abs/2410.03966} {arXiv:2410.03966 [hep-ph]} \BibitemShut
  {NoStop}%
\bibitem [{\citenamefont {Ji}(2013)}]{Ji:2013dva}%
  \BibitemOpen
  \bibfield  {author} {\bibinfo {author} {\bibfnamefont {X.}~\bibnamefont
  {Ji}},\ }\bibfield  {title} {\bibinfo {title} {{Parton Physics on a Euclidean
  Lattice}},\ }\href {https://doi.org/10.1103/PhysRevLett.110.262002}
  {\bibfield  {journal} {\bibinfo  {journal} {Phys. Rev. Lett.}\ }\textbf
  {\bibinfo {volume} {110}},\ \bibinfo {pages} {262002} (\bibinfo {year}
  {2013})},\ \Eprint {https://arxiv.org/abs/1305.1539} {arXiv:1305.1539
  [hep-ph]} \BibitemShut {NoStop}%
\bibitem [{\citenamefont {Ji}(2014)}]{Ji:2014gla}%
  \BibitemOpen
  \bibfield  {author} {\bibinfo {author} {\bibfnamefont {X.}~\bibnamefont
  {Ji}},\ }\bibfield  {title} {\bibinfo {title} {{Parton Physics from
  Large-Momentum Effective Field Theory}},\ }\href
  {https://doi.org/10.1007/s11433-014-5492-3} {\bibfield  {journal} {\bibinfo
  {journal} {Sci. China Phys. Mech. Astron.}\ }\textbf {\bibinfo {volume}
  {57}},\ \bibinfo {pages} {1407} (\bibinfo {year} {2014})},\ \Eprint
  {https://arxiv.org/abs/1404.6680} {arXiv:1404.6680 [hep-ph]} \BibitemShut
  {NoStop}%
\bibitem [{\citenamefont {Ji}\ \emph {et~al.}(2021)\citenamefont {Ji},
  \citenamefont {Liu}, \citenamefont {Liu}, \citenamefont {Zhang},\ and\
  \citenamefont {Zhao}}]{Ji:2020ect}%
  \BibitemOpen
  \bibfield  {author} {\bibinfo {author} {\bibfnamefont {X.}~\bibnamefont
  {Ji}}, \bibinfo {author} {\bibfnamefont {Y.-S.}\ \bibnamefont {Liu}},
  \bibinfo {author} {\bibfnamefont {Y.}~\bibnamefont {Liu}}, \bibinfo {author}
  {\bibfnamefont {J.-H.}\ \bibnamefont {Zhang}},\ and\ \bibinfo {author}
  {\bibfnamefont {Y.}~\bibnamefont {Zhao}},\ }\bibfield  {title} {\bibinfo
  {title} {{Large-momentum effective theory}},\ }\href
  {https://doi.org/10.1103/RevModPhys.93.035005} {\bibfield  {journal}
  {\bibinfo  {journal} {Rev. Mod. Phys.}\ }\textbf {\bibinfo {volume} {93}},\
  \bibinfo {pages} {035005} (\bibinfo {year} {2021})},\ \Eprint
  {https://arxiv.org/abs/2004.03543} {arXiv:2004.03543 [hep-ph]} \BibitemShut
  {NoStop}%
\bibitem [{\citenamefont {Zhang}\ \emph {et~al.}(2017)\citenamefont {Zhang},
  \citenamefont {Chen}, \citenamefont {Ji}, \citenamefont {Jin},\ and\
  \citenamefont {Lin}}]{Zhang:2017bzy}%
  \BibitemOpen
  \bibfield  {author} {\bibinfo {author} {\bibfnamefont {J.-H.}\ \bibnamefont
  {Zhang}}, \bibinfo {author} {\bibfnamefont {J.-W.}\ \bibnamefont {Chen}},
  \bibinfo {author} {\bibfnamefont {X.}~\bibnamefont {Ji}}, \bibinfo {author}
  {\bibfnamefont {L.}~\bibnamefont {Jin}},\ and\ \bibinfo {author}
  {\bibfnamefont {H.-W.}\ \bibnamefont {Lin}},\ }\bibfield  {title} {\bibinfo
  {title} {{Pion Distribution Amplitude from Lattice QCD}},\ }\href
  {https://doi.org/10.1103/PhysRevD.95.094514} {\bibfield  {journal} {\bibinfo
  {journal} {Phys. Rev. D}\ }\textbf {\bibinfo {volume} {95}},\ \bibinfo
  {pages} {094514} (\bibinfo {year} {2017})},\ \Eprint
  {https://arxiv.org/abs/1702.00008} {arXiv:1702.00008 [hep-lat]} \BibitemShut
  {NoStop}%
\bibitem [{\citenamefont {Hua}\ \emph {et~al.}(2022)\citenamefont {Hua} \emph
  {et~al.}}]{LatticeParton:2022zqc}%
  \BibitemOpen
  \bibfield  {author} {\bibinfo {author} {\bibfnamefont {J.}~\bibnamefont
  {Hua}} \emph {et~al.} (\bibinfo {collaboration} {Lattice Parton}),\
  }\bibfield  {title} {\bibinfo {title} {{Pion and Kaon Distribution Amplitudes
  from Lattice QCD}},\ }\href {https://doi.org/10.1103/PhysRevLett.129.132001}
  {\bibfield  {journal} {\bibinfo  {journal} {Phys. Rev. Lett.}\ }\textbf
  {\bibinfo {volume} {129}},\ \bibinfo {pages} {132001} (\bibinfo {year}
  {2022})},\ \Eprint {https://arxiv.org/abs/2201.09173} {arXiv:2201.09173
  [hep-lat]} \BibitemShut {NoStop}%
\bibitem [{\citenamefont {Holligan}\ \emph {et~al.}(2023)\citenamefont
  {Holligan}, \citenamefont {Ji}, \citenamefont {Lin}, \citenamefont {Su},\
  and\ \citenamefont {Zhang}}]{Holligan:2023rex}%
  \BibitemOpen
  \bibfield  {author} {\bibinfo {author} {\bibfnamefont {J.}~\bibnamefont
  {Holligan}}, \bibinfo {author} {\bibfnamefont {X.}~\bibnamefont {Ji}},
  \bibinfo {author} {\bibfnamefont {H.-W.}\ \bibnamefont {Lin}}, \bibinfo
  {author} {\bibfnamefont {Y.}~\bibnamefont {Su}},\ and\ \bibinfo {author}
  {\bibfnamefont {R.}~\bibnamefont {Zhang}},\ }\bibfield  {title} {\bibinfo
  {title} {{Precision control in lattice calculation of x-dependent pion
  distribution amplitude}},\ }\href
  {https://doi.org/10.1016/j.nuclphysb.2023.116282} {\bibfield  {journal}
  {\bibinfo  {journal} {Nucl. Phys. B}\ }\textbf {\bibinfo {volume} {993}},\
  \bibinfo {pages} {116282} (\bibinfo {year} {2023})},\ \Eprint
  {https://arxiv.org/abs/2301.10372} {arXiv:2301.10372 [hep-lat]} \BibitemShut
  {NoStop}%
\bibitem [{\citenamefont {Chu}\ \emph {et~al.}(2024)\citenamefont {Chu} \emph
  {et~al.}}]{LatticeParton:2023xdl}%
  \BibitemOpen
  \bibfield  {author} {\bibinfo {author} {\bibfnamefont {M.-H.}\ \bibnamefont
  {Chu}} \emph {et~al.} (\bibinfo {collaboration} {Lattice Parton}),\
  }\bibfield  {title} {\bibinfo {title} {{Transverse-momentum-dependent wave
  functions of the pion from lattice QCD}},\ }\href
  {https://doi.org/10.1103/PhysRevD.109.L091503} {\bibfield  {journal}
  {\bibinfo  {journal} {Phys. Rev. D}\ }\textbf {\bibinfo {volume} {109}},\
  \bibinfo {pages} {L091503} (\bibinfo {year} {2024})},\ \Eprint
  {https://arxiv.org/abs/2302.09961} {arXiv:2302.09961 [hep-lat]} \BibitemShut
  {NoStop}%
\bibitem [{\citenamefont {Mitter}\ \emph {et~al.}(2015)\citenamefont {Mitter},
  \citenamefont {Pawlowski},\ and\ \citenamefont
  {Strodthoff}}]{Mitter:2014wpa}%
  \BibitemOpen
  \bibfield  {author} {\bibinfo {author} {\bibfnamefont {M.}~\bibnamefont
  {Mitter}}, \bibinfo {author} {\bibfnamefont {J.~M.}\ \bibnamefont
  {Pawlowski}},\ and\ \bibinfo {author} {\bibfnamefont {N.}~\bibnamefont
  {Strodthoff}},\ }\bibfield  {title} {\bibinfo {title} {{Chiral symmetry
  breaking in continuum QCD}},\ }\href
  {https://doi.org/10.1103/PhysRevD.91.054035} {\bibfield  {journal} {\bibinfo
  {journal} {Phys. Rev.}\ }\textbf {\bibinfo {volume} {D91}},\ \bibinfo {pages}
  {054035} (\bibinfo {year} {2015})},\ \Eprint
  {https://arxiv.org/abs/1411.7978} {arXiv:1411.7978 [hep-ph]} \BibitemShut
  {NoStop}%
\bibitem [{\citenamefont {Braun}\ \emph {et~al.}(2016)\citenamefont {Braun},
  \citenamefont {Fister}, \citenamefont {Pawlowski},\ and\ \citenamefont
  {Rennecke}}]{Braun:2014ata}%
  \BibitemOpen
  \bibfield  {author} {\bibinfo {author} {\bibfnamefont {J.}~\bibnamefont
  {Braun}}, \bibinfo {author} {\bibfnamefont {L.}~\bibnamefont {Fister}},
  \bibinfo {author} {\bibfnamefont {J.~M.}\ \bibnamefont {Pawlowski}},\ and\
  \bibinfo {author} {\bibfnamefont {F.}~\bibnamefont {Rennecke}},\ }\bibfield
  {title} {\bibinfo {title} {{From Quarks and Gluons to Hadrons: Chiral
  Symmetry Breaking in Dynamical QCD}},\ }\href
  {https://doi.org/10.1103/PhysRevD.94.034016} {\bibfield  {journal} {\bibinfo
  {journal} {Phys. Rev.}\ }\textbf {\bibinfo {volume} {D94}},\ \bibinfo {pages}
  {034016} (\bibinfo {year} {2016})},\ \Eprint
  {https://arxiv.org/abs/1412.1045} {arXiv:1412.1045 [hep-ph]} \BibitemShut
  {NoStop}%
\bibitem [{\citenamefont {Rennecke}(2015)}]{Rennecke:2015eba}%
  \BibitemOpen
  \bibfield  {author} {\bibinfo {author} {\bibfnamefont {F.}~\bibnamefont
  {Rennecke}},\ }\bibfield  {title} {\bibinfo {title} {{Vacuum structure of
  vector mesons in QCD}},\ }\href {https://doi.org/10.1103/PhysRevD.92.076012}
  {\bibfield  {journal} {\bibinfo  {journal} {Phys. Rev.}\ }\textbf {\bibinfo
  {volume} {D92}},\ \bibinfo {pages} {076012} (\bibinfo {year} {2015})},\
  \Eprint {https://arxiv.org/abs/1504.03585} {arXiv:1504.03585 [hep-ph]}
  \BibitemShut {NoStop}%
\bibitem [{\citenamefont {Cyrol}\ \emph {et~al.}(2016)\citenamefont {Cyrol},
  \citenamefont {Fister}, \citenamefont {Mitter}, \citenamefont {Pawlowski},\
  and\ \citenamefont {Strodthoff}}]{Cyrol:2016tym}%
  \BibitemOpen
  \bibfield  {author} {\bibinfo {author} {\bibfnamefont {A.~K.}\ \bibnamefont
  {Cyrol}}, \bibinfo {author} {\bibfnamefont {L.}~\bibnamefont {Fister}},
  \bibinfo {author} {\bibfnamefont {M.}~\bibnamefont {Mitter}}, \bibinfo
  {author} {\bibfnamefont {J.~M.}\ \bibnamefont {Pawlowski}},\ and\ \bibinfo
  {author} {\bibfnamefont {N.}~\bibnamefont {Strodthoff}},\ }\bibfield  {title}
  {\bibinfo {title} {{Landau gauge Yang-Mills correlation functions}},\ }\href
  {https://doi.org/10.1103/PhysRevD.94.054005} {\bibfield  {journal} {\bibinfo
  {journal} {Phys. Rev.}\ }\textbf {\bibinfo {volume} {D94}},\ \bibinfo {pages}
  {054005} (\bibinfo {year} {2016})},\ \Eprint
  {https://arxiv.org/abs/1605.01856} {arXiv:1605.01856 [hep-ph]} \BibitemShut
  {NoStop}%
\bibitem [{\citenamefont {Cyrol}\ \emph {et~al.}(2018)\citenamefont {Cyrol},
  \citenamefont {Mitter}, \citenamefont {Pawlowski},\ and\ \citenamefont
  {Strodthoff}}]{Cyrol:2017ewj}%
  \BibitemOpen
  \bibfield  {author} {\bibinfo {author} {\bibfnamefont {A.~K.}\ \bibnamefont
  {Cyrol}}, \bibinfo {author} {\bibfnamefont {M.}~\bibnamefont {Mitter}},
  \bibinfo {author} {\bibfnamefont {J.~M.}\ \bibnamefont {Pawlowski}},\ and\
  \bibinfo {author} {\bibfnamefont {N.}~\bibnamefont {Strodthoff}},\ }\bibfield
   {title} {\bibinfo {title} {{Nonperturbative quark, gluon, and meson
  correlators of unquenched QCD}},\ }\href
  {https://doi.org/10.1103/PhysRevD.97.054006} {\bibfield  {journal} {\bibinfo
  {journal} {Phys. Rev.}\ }\textbf {\bibinfo {volume} {D97}},\ \bibinfo {pages}
  {054006} (\bibinfo {year} {2018})},\ \Eprint
  {https://arxiv.org/abs/1706.06326} {arXiv:1706.06326 [hep-ph]} \BibitemShut
  {NoStop}%
\bibitem [{\citenamefont {Corell}\ \emph {et~al.}(2018)\citenamefont {Corell},
  \citenamefont {Cyrol}, \citenamefont {Mitter}, \citenamefont {Pawlowski},\
  and\ \citenamefont {Strodthoff}}]{Corell:2018yil}%
  \BibitemOpen
  \bibfield  {author} {\bibinfo {author} {\bibfnamefont {L.}~\bibnamefont
  {Corell}}, \bibinfo {author} {\bibfnamefont {A.~K.}\ \bibnamefont {Cyrol}},
  \bibinfo {author} {\bibfnamefont {M.}~\bibnamefont {Mitter}}, \bibinfo
  {author} {\bibfnamefont {J.~M.}\ \bibnamefont {Pawlowski}},\ and\ \bibinfo
  {author} {\bibfnamefont {N.}~\bibnamefont {Strodthoff}},\ }\bibfield  {title}
  {\bibinfo {title} {{Correlation functions of three-dimensional Yang-Mills
  theory from the FRG}},\ }\href {https://doi.org/10.21468/SciPostPhys.5.6.066}
  {\bibfield  {journal} {\bibinfo  {journal} {SciPost Phys.}\ }\textbf
  {\bibinfo {volume} {5}},\ \bibinfo {pages} {066} (\bibinfo {year} {2018})},\
  \Eprint {https://arxiv.org/abs/1803.10092} {arXiv:1803.10092 [hep-ph]}
  \BibitemShut {NoStop}%
\bibitem [{\citenamefont {Fu}\ \emph {et~al.}(2020)\citenamefont {Fu},
  \citenamefont {Pawlowski},\ and\ \citenamefont {Rennecke}}]{Fu:2019hdw}%
  \BibitemOpen
  \bibfield  {author} {\bibinfo {author} {\bibfnamefont {W.-j.}\ \bibnamefont
  {Fu}}, \bibinfo {author} {\bibfnamefont {J.~M.}\ \bibnamefont {Pawlowski}},\
  and\ \bibinfo {author} {\bibfnamefont {F.}~\bibnamefont {Rennecke}},\
  }\bibfield  {title} {\bibinfo {title} {{QCD phase structure at finite
  temperature and density}},\ }\href
  {https://doi.org/10.1103/PhysRevD.101.054032} {\bibfield  {journal} {\bibinfo
   {journal} {Phys. Rev. D}\ }\textbf {\bibinfo {volume} {101}},\ \bibinfo
  {pages} {054032} (\bibinfo {year} {2020})},\ \Eprint
  {https://arxiv.org/abs/1909.02991} {arXiv:1909.02991 [hep-ph]} \BibitemShut
  {NoStop}%
\bibitem [{\citenamefont {Braun}\ \emph {et~al.}(2020)\citenamefont {Braun},
  \citenamefont {Fu}, \citenamefont {Pawlowski}, \citenamefont {Rennecke},
  \citenamefont {Rosenbl\"uh},\ and\ \citenamefont {Yin}}]{Braun:2020ada}%
  \BibitemOpen
  \bibfield  {author} {\bibinfo {author} {\bibfnamefont {J.}~\bibnamefont
  {Braun}}, \bibinfo {author} {\bibfnamefont {W.-j.}\ \bibnamefont {Fu}},
  \bibinfo {author} {\bibfnamefont {J.~M.}\ \bibnamefont {Pawlowski}}, \bibinfo
  {author} {\bibfnamefont {F.}~\bibnamefont {Rennecke}}, \bibinfo {author}
  {\bibfnamefont {D.}~\bibnamefont {Rosenbl\"uh}},\ and\ \bibinfo {author}
  {\bibfnamefont {S.}~\bibnamefont {Yin}},\ }\bibfield  {title} {\bibinfo
  {title} {{Chiral susceptibility in ( 2+1 )-flavor QCD}},\ }\href
  {https://doi.org/10.1103/PhysRevD.102.056010} {\bibfield  {journal} {\bibinfo
   {journal} {Phys. Rev. D}\ }\textbf {\bibinfo {volume} {102}},\ \bibinfo
  {pages} {056010} (\bibinfo {year} {2020})},\ \Eprint
  {https://arxiv.org/abs/2003.13112} {arXiv:2003.13112 [hep-ph]} \BibitemShut
  {NoStop}%
\bibitem [{\citenamefont {Braun}\ \emph {et~al.}(2023)\citenamefont {Braun}
  \emph {et~al.}}]{Braun:2023qak}%
  \BibitemOpen
  \bibfield  {author} {\bibinfo {author} {\bibfnamefont {J.}~\bibnamefont
  {Braun}} \emph {et~al.},\ }\bibfield  {title} {\bibinfo {title} {{Soft modes
  in hot QCD matter}},\ }\href@noop {} {\  (\bibinfo {year} {2023})},\ \Eprint
  {https://arxiv.org/abs/2310.19853} {arXiv:2310.19853 [hep-ph]} \BibitemShut
  {NoStop}%
\bibitem [{\citenamefont {Ihssen}\ \emph {et~al.}(2024)\citenamefont {Ihssen},
  \citenamefont {Pawlowski}, \citenamefont {Sattler},\ and\ \citenamefont
  {Wink}}]{Ihssen:2024miv}%
  \BibitemOpen
  \bibfield  {author} {\bibinfo {author} {\bibfnamefont {F.}~\bibnamefont
  {Ihssen}}, \bibinfo {author} {\bibfnamefont {J.~M.}\ \bibnamefont
  {Pawlowski}}, \bibinfo {author} {\bibfnamefont {F.~R.}\ \bibnamefont
  {Sattler}},\ and\ \bibinfo {author} {\bibfnamefont {N.}~\bibnamefont
  {Wink}},\ }\bibfield  {title} {\bibinfo {title} {{Towards quantitative
  precision in functional QCD I}},\ }\href@noop {} {\  (\bibinfo {year}
  {2024})},\ \Eprint {https://arxiv.org/abs/2408.08413} {arXiv:2408.08413
  [hep-ph]} \BibitemShut {NoStop}%
\bibitem [{\citenamefont {Fu}\ \emph {et~al.}(2024{\natexlab{a}})\citenamefont
  {Fu}, \citenamefont {Pawlowski}, \citenamefont {Pisarski}, \citenamefont
  {Rennecke}, \citenamefont {Wen},\ and\ \citenamefont {Yin}}]{Fu:2024rto}%
  \BibitemOpen
  \bibfield  {author} {\bibinfo {author} {\bibfnamefont {W.-j.}\ \bibnamefont
  {Fu}}, \bibinfo {author} {\bibfnamefont {J.~M.}\ \bibnamefont {Pawlowski}},
  \bibinfo {author} {\bibfnamefont {R.~D.}\ \bibnamefont {Pisarski}}, \bibinfo
  {author} {\bibfnamefont {F.}~\bibnamefont {Rennecke}}, \bibinfo {author}
  {\bibfnamefont {R.}~\bibnamefont {Wen}},\ and\ \bibinfo {author}
  {\bibfnamefont {S.}~\bibnamefont {Yin}},\ }\bibfield  {title} {\bibinfo
  {title} {{The QCD moat regime and its real-time properties}},\ }\href@noop {}
  {\  (\bibinfo {year} {2024}{\natexlab{a}})},\ \Eprint
  {https://arxiv.org/abs/2412.15949} {arXiv:2412.15949 [hep-ph]} \BibitemShut
  {NoStop}%
\bibitem [{\citenamefont {Dupuis}\ \emph {et~al.}(2021)\citenamefont {Dupuis},
  \citenamefont {Canet}, \citenamefont {Eichhorn}, \citenamefont {Metzner},
  \citenamefont {Pawlowski}, \citenamefont {Tissier},\ and\ \citenamefont
  {Wschebor}}]{Dupuis:2020fhh}%
  \BibitemOpen
  \bibfield  {author} {\bibinfo {author} {\bibfnamefont {N.}~\bibnamefont
  {Dupuis}}, \bibinfo {author} {\bibfnamefont {L.}~\bibnamefont {Canet}},
  \bibinfo {author} {\bibfnamefont {A.}~\bibnamefont {Eichhorn}}, \bibinfo
  {author} {\bibfnamefont {W.}~\bibnamefont {Metzner}}, \bibinfo {author}
  {\bibfnamefont {J.~M.}\ \bibnamefont {Pawlowski}}, \bibinfo {author}
  {\bibfnamefont {M.}~\bibnamefont {Tissier}},\ and\ \bibinfo {author}
  {\bibfnamefont {N.}~\bibnamefont {Wschebor}},\ }\bibfield  {title} {\bibinfo
  {title} {{The nonperturbative functional renormalization group and its
  applications}},\ }\href {https://doi.org/10.1016/j.physrep.2021.01.001}
  {\bibfield  {journal} {\bibinfo  {journal} {Phys. Rept.}\ }\textbf {\bibinfo
  {volume} {910}},\ \bibinfo {pages} {1} (\bibinfo {year} {2021})},\ \Eprint
  {https://arxiv.org/abs/2006.04853} {arXiv:2006.04853 [cond-mat.stat-mech]}
  \BibitemShut {NoStop}%
\bibitem [{\citenamefont {Fu}(2022)}]{Fu:2022gou}%
  \BibitemOpen
  \bibfield  {author} {\bibinfo {author} {\bibfnamefont {W.-j.}\ \bibnamefont
  {Fu}},\ }\bibfield  {title} {\bibinfo {title} {{QCD at finite temperature and
  density within the fRG approach: an overview}},\ }\href
  {https://doi.org/10.1088/1572-9494/ac86be} {\bibfield  {journal} {\bibinfo
  {journal} {Commun. Theor. Phys.}\ }\textbf {\bibinfo {volume} {74}},\
  \bibinfo {pages} {097304} (\bibinfo {year} {2022})},\ \Eprint
  {https://arxiv.org/abs/2205.00468} {arXiv:2205.00468 [hep-ph]} \BibitemShut
  {NoStop}%
\bibitem [{\citenamefont {Fu}\ \emph {et~al.}(2023)\citenamefont {Fu},
  \citenamefont {Huang}, \citenamefont {Pawlowski},\ and\ \citenamefont
  {Tan}}]{Fu:2022uow}%
  \BibitemOpen
  \bibfield  {author} {\bibinfo {author} {\bibfnamefont {W.-j.}\ \bibnamefont
  {Fu}}, \bibinfo {author} {\bibfnamefont {C.}~\bibnamefont {Huang}}, \bibinfo
  {author} {\bibfnamefont {J.~M.}\ \bibnamefont {Pawlowski}},\ and\ \bibinfo
  {author} {\bibfnamefont {Y.-y.}\ \bibnamefont {Tan}},\ }\bibfield  {title}
  {\bibinfo {title} {{Four-quark scatterings in QCD I}},\ }\href
  {https://doi.org/10.21468/SciPostPhys.14.4.069} {\bibfield  {journal}
  {\bibinfo  {journal} {SciPost Phys.}\ }\textbf {\bibinfo {volume} {14}},\
  \bibinfo {pages} {069} (\bibinfo {year} {2023})},\ \Eprint
  {https://arxiv.org/abs/2209.13120} {arXiv:2209.13120 [hep-ph]} \BibitemShut
  {NoStop}%
\bibitem [{\citenamefont {Fu}\ \emph {et~al.}(2024{\natexlab{b}})\citenamefont
  {Fu}, \citenamefont {Huang}, \citenamefont {Pawlowski},\ and\ \citenamefont
  {Tan}}]{Fu:2024ysj}%
  \BibitemOpen
  \bibfield  {author} {\bibinfo {author} {\bibfnamefont {W.-j.}\ \bibnamefont
  {Fu}}, \bibinfo {author} {\bibfnamefont {C.}~\bibnamefont {Huang}}, \bibinfo
  {author} {\bibfnamefont {J.~M.}\ \bibnamefont {Pawlowski}},\ and\ \bibinfo
  {author} {\bibfnamefont {Y.-y.}\ \bibnamefont {Tan}},\ }\bibfield  {title}
  {\bibinfo {title} {{Four-quark scatterings in QCD II}},\ }\href
  {https://doi.org/10.21468/SciPostPhys.17.5.148} {\bibfield  {journal}
  {\bibinfo  {journal} {SciPost Phys.}\ }\textbf {\bibinfo {volume} {17}},\
  \bibinfo {pages} {148} (\bibinfo {year} {2024}{\natexlab{b}})},\ \Eprint
  {https://arxiv.org/abs/2401.07638} {arXiv:2401.07638 [hep-ph]} \BibitemShut
  {NoStop}%
\bibitem [{\citenamefont {Fu}\ \emph {et~al.}(2025)\citenamefont {Fu},
  \citenamefont {Huang}, \citenamefont {Pawlowski}, \citenamefont {Tan},\ and\
  \citenamefont {Zhou}}]{Fu:2025hcm}%
  \BibitemOpen
  \bibfield  {author} {\bibinfo {author} {\bibfnamefont {W.-j.}\ \bibnamefont
  {Fu}}, \bibinfo {author} {\bibfnamefont {C.}~\bibnamefont {Huang}}, \bibinfo
  {author} {\bibfnamefont {J.~M.}\ \bibnamefont {Pawlowski}}, \bibinfo {author}
  {\bibfnamefont {Y.-y.}\ \bibnamefont {Tan}},\ and\ \bibinfo {author}
  {\bibfnamefont {L.-j.}\ \bibnamefont {Zhou}},\ }\bibfield  {title} {\bibinfo
  {title} {{Four-quark scatterings in QCD III}},\ }\href@noop {} {\  (\bibinfo
  {year} {2025})},\ \Eprint {https://arxiv.org/abs/2502.14388}
  {arXiv:2502.14388 [hep-ph]} \BibitemShut {NoStop}%
\bibitem [{\citenamefont {Xu}\ \emph {et~al.}(2018)\citenamefont {Xu},
  \citenamefont {Chang}, \citenamefont {Roberts},\ and\ \citenamefont
  {Zong}}]{Xu:2018eii}%
  \BibitemOpen
  \bibfield  {author} {\bibinfo {author} {\bibfnamefont {S.-S.}\ \bibnamefont
  {Xu}}, \bibinfo {author} {\bibfnamefont {L.}~\bibnamefont {Chang}}, \bibinfo
  {author} {\bibfnamefont {C.~D.}\ \bibnamefont {Roberts}},\ and\ \bibinfo
  {author} {\bibfnamefont {H.-S.}\ \bibnamefont {Zong}},\ }\bibfield  {title}
  {\bibinfo {title} {{Pion and kaon valence-quark parton quasidistributions}},\
  }\href {https://doi.org/10.1103/PhysRevD.97.094014} {\bibfield  {journal}
  {\bibinfo  {journal} {Phys. Rev. D}\ }\textbf {\bibinfo {volume} {97}},\
  \bibinfo {pages} {094014} (\bibinfo {year} {2018})},\ \Eprint
  {https://arxiv.org/abs/1802.09552} {arXiv:1802.09552 [nucl-th]} \BibitemShut
  {NoStop}%
\bibitem [{\citenamefont {Wetterich}(1993)}]{Wetterich:1992yh}%
  \BibitemOpen
  \bibfield  {author} {\bibinfo {author} {\bibfnamefont {C.}~\bibnamefont
  {Wetterich}},\ }\bibfield  {title} {\bibinfo {title} {{Exact evolution
  equation for the effective potential}},\ }\href
  {https://doi.org/10.1016/0370-2693(93)90726-X} {\bibfield  {journal}
  {\bibinfo  {journal} {Phys. Lett.}\ }\textbf {\bibinfo {volume} {B301}},\
  \bibinfo {pages} {90} (\bibinfo {year} {1993})}\BibitemShut {NoStop}%
\bibitem [{\citenamefont {Roberts}\ \emph {et~al.}(2021)\citenamefont
  {Roberts}, \citenamefont {Richards}, \citenamefont {Horn},\ and\
  \citenamefont {Chang}}]{Roberts:2021nhw}%
  \BibitemOpen
  \bibfield  {author} {\bibinfo {author} {\bibfnamefont {C.~D.}\ \bibnamefont
  {Roberts}}, \bibinfo {author} {\bibfnamefont {D.~G.}\ \bibnamefont
  {Richards}}, \bibinfo {author} {\bibfnamefont {T.}~\bibnamefont {Horn}},\
  and\ \bibinfo {author} {\bibfnamefont {L.}~\bibnamefont {Chang}},\ }\bibfield
   {title} {\bibinfo {title} {{Insights into the emergence of mass from studies
  of pion and kaon structure}},\ }\href
  {https://doi.org/10.1016/j.ppnp.2021.103883} {\bibfield  {journal} {\bibinfo
  {journal} {Prog. Part. Nucl. Phys.}\ }\textbf {\bibinfo {volume} {120}},\
  \bibinfo {pages} {103883} (\bibinfo {year} {2021})},\ \Eprint
  {https://arxiv.org/abs/2102.01765} {arXiv:2102.01765 [hep-ph]} \BibitemShut
  {NoStop}%
\bibitem [{\citenamefont {Ball}\ \emph {et~al.}(2007)\citenamefont {Ball},
  \citenamefont {Braun},\ and\ \citenamefont {Lenz}}]{Ball:2007zt}%
  \BibitemOpen
  \bibfield  {author} {\bibinfo {author} {\bibfnamefont {P.}~\bibnamefont
  {Ball}}, \bibinfo {author} {\bibfnamefont {V.~M.}\ \bibnamefont {Braun}},\
  and\ \bibinfo {author} {\bibfnamefont {A.}~\bibnamefont {Lenz}},\ }\bibfield
  {title} {\bibinfo {title} {{Twist-4 distribution amplitudes of the K* and phi
  mesons in QCD}},\ }\href {https://doi.org/10.1088/1126-6708/2007/08/090}
  {\bibfield  {journal} {\bibinfo  {journal} {JHEP}\ }\textbf {\bibinfo
  {volume} {08}},\ \bibinfo {pages} {090}},\ \Eprint
  {https://arxiv.org/abs/0707.1201} {arXiv:0707.1201 [hep-ph]} \BibitemShut
  {NoStop}%
\bibitem [{\citenamefont {Zhong}\ \emph {et~al.}(2021)\citenamefont {Zhong},
  \citenamefont {Zhu}, \citenamefont {Fu}, \citenamefont {Wu},\ and\
  \citenamefont {Huang}}]{Zhong:2021epq}%
  \BibitemOpen
  \bibfield  {author} {\bibinfo {author} {\bibfnamefont {T.}~\bibnamefont
  {Zhong}}, \bibinfo {author} {\bibfnamefont {Z.-H.}\ \bibnamefont {Zhu}},
  \bibinfo {author} {\bibfnamefont {H.-B.}\ \bibnamefont {Fu}}, \bibinfo
  {author} {\bibfnamefont {X.-G.}\ \bibnamefont {Wu}},\ and\ \bibinfo {author}
  {\bibfnamefont {T.}~\bibnamefont {Huang}},\ }\bibfield  {title} {\bibinfo
  {title} {{Improved light-cone harmonic oscillator model for the pionic
  leading-twist distribution amplitude}},\ }\href
  {https://doi.org/10.1103/PhysRevD.104.016021} {\bibfield  {journal} {\bibinfo
   {journal} {Phys. Rev. D}\ }\textbf {\bibinfo {volume} {104}},\ \bibinfo
  {pages} {016021} (\bibinfo {year} {2021})},\ \Eprint
  {https://arxiv.org/abs/2102.03989} {arXiv:2102.03989 [hep-ph]} \BibitemShut
  {NoStop}%
\bibitem [{\citenamefont {Roberts}\ \emph {et~al.}(2010)\citenamefont
  {Roberts}, \citenamefont {Roberts}, \citenamefont {Bashir}, \citenamefont
  {Gutierrez-Guerrero},\ and\ \citenamefont {Tandy}}]{Roberts:2010rn}%
  \BibitemOpen
  \bibfield  {author} {\bibinfo {author} {\bibfnamefont {H.~L.~L.}\
  \bibnamefont {Roberts}}, \bibinfo {author} {\bibfnamefont {C.~D.}\
  \bibnamefont {Roberts}}, \bibinfo {author} {\bibfnamefont {A.}~\bibnamefont
  {Bashir}}, \bibinfo {author} {\bibfnamefont {L.~X.}\ \bibnamefont
  {Gutierrez-Guerrero}},\ and\ \bibinfo {author} {\bibfnamefont {P.~C.}\
  \bibnamefont {Tandy}},\ }\bibfield  {title} {\bibinfo {title} {{Abelian
  anomaly and neutral pion production}},\ }\href
  {https://doi.org/10.1103/PhysRevC.82.065202} {\bibfield  {journal} {\bibinfo
  {journal} {Phys. Rev. C}\ }\textbf {\bibinfo {volume} {82}},\ \bibinfo
  {pages} {065202} (\bibinfo {year} {2010})},\ \Eprint
  {https://arxiv.org/abs/1009.0067} {arXiv:1009.0067 [nucl-th]} \BibitemShut
  {NoStop}%
\end{thebibliography}%

\end{document}